\documentclass[superscriptaddress, reprint]{revtex4-1}
\usepackage{hyperref}
\usepackage[version=4]{mhchem} %chemical formulae
\usepackage{graphicx}
\usepackage{chemmacros}
\usepackage{gensymb}
\usepackage{dcolumn}
\usepackage{amssymb}

{\begin{document}

%title 
\title{Unusual effects of magnetic dilution in the ferrimagnetic columnar ordered \ce{Sm2MnMnMn_{4-x}Ti_xO_{12}} perovskites}

%authors 
\author{Anuradha M. Vibhakar}
\affiliation{Clarendon Laboratory, Department of Physics, University of Oxford, Oxford, OX1 3PU, United Kingdom}
\affiliation{Department of Physics and Astronomy, University College London, Gower Street, London, WC1E 6BT, United Kingdom}
\author{Dmitry D. Khalyavin}
\affiliation{ ISIS facility, Rutherford Appleton Laboratory-STFC, Chilton, Didcot, OX11 0QX, United Kingdom}
\author{Pascal Manuel}
\affiliation{ ISIS facility, Rutherford Appleton Laboratory-STFC, Chilton, Didcot, OX11 0QX, United Kingdom}
\author{Ran Liu}
\affiliation{International Center for Materials Nanoarchitectonics (WPI-MANA), National Institute for Materials Science (NIMS), Namiki 1-1, Tsukuba, Ibaraki 305-0044, Japan}
\affiliation{Graduate School of Chemical Sciences and Engineering, Hokkaido University, North 10 West 8, Kita-ku, Sapporo, Hokkaido 060-0810, Japan}
\author{Kazunari Yamaura}
\affiliation{International Center for Materials Nanoarchitectonics (WPI-MANA), National Institute for Materials Science (NIMS), Namiki 1-1, Tsukuba, Ibaraki 305-0044, Japan}
\affiliation{Graduate School of Chemical Sciences and Engineering, Hokkaido University, North 10 West 8, Kita-ku, Sapporo, Hokkaido 060-0810, Japan}
\author{Alexei A. Belik}
\affiliation{International Center for Materials Nanoarchitectonics (WPI-MANA), National Institute for Materials Science (NIMS), Namiki 1-1, Tsukuba, Ibaraki 305-0044, Japan}
\author{Roger D. Johnson}
\affiliation{Clarendon Laboratory, Department of Physics, University of Oxford, Oxford, OX1 3PU, United Kingdom}
\affiliation{Department of Physics and Astronomy, University College London, Gower Street, London, WC1E 6BT, United Kingdom}

\date{\today}

\begin{abstract}
Powder neutron diffraction experiments have been employed to establish the effects of site-selective magnetic dilution in the \ce{Sm2MnMnMn_{4-x}Ti_xO_{12}} A-site columnar ordered quadruple perovskite manganites ($x=1$, $x=2$ and $x=3$). We show that in all three compositions the Mn ions adopt a collinear ferrimagnetic structure below 27 K, 62 K and 34 K, respectively. An unexpected increase in the ordering temperature was observed between the $x=1$ and $x=2$ samples, which indicates a considerable departure from mean field behaviour. This result is corroborated by large reductions in the theoretical ground state magnetic moments observed across the series, which indicate the presence of spin fluctuations and or disorder. We show that long range magnetic order in the $x = 3$ sample, which occurs below the percolation threshold for B-B exchange, can only be understood to arise if magnetic order in \ce{Sm2MnMnMn_{4-x}Ti_xO_{12}} is mediated via \emph{both} A-B and B-B exchange, hence confirming the importance of A-B exchange interactions in these materials. Finally we show that site-selective magnetic dilution enables the tuning of a ferrimagnetic compensation point and the introduction of temperature-induced magnetization reversal.
\end{abstract}
 
 %isodistort a1 e t
 %check numerical errors 
\maketitle
 
\section{Introduction}\label{SEC::introduction}

The simple perovskite manganites (general chemical formula \ce{ABO3}, B=Mn) are canonical examples of correlated electron systems in which charge, spin and orbital order can be tuned to give rise to exotic electronic ground states. This is most famously demonstrated in \ce{La_{1-x}A$'$_xMnO3} (A$'$ = Ca$^{2+}$, Sr$^{2+}$, and Ba$^{2+}$) where a transition from an antiferromagnetic insulating phase to a ferromagnetic metallic phase, driven by dopant concentration, leads to the emergence of the technologically important property of colossal magnetoresistance \cite{1994Jin}. More broadly, when the A-site is occupied by a rare-earth ion or yttrium (R) a variety of B-site magnetic structures have been found. For example, A-type antiferromagnetic order was observed experimentally for R = La {\textrightarrow} Gd \cite{1955Wollan, 2006Zhou}, and for compositions with smaller R$^{3+}$ ionic radii (Ho {\textrightarrow} Yb) E-type antiferromagnetic order is stabilised \cite{2006Zhou}. Furthermore, in the mixed valence manganites, such as \ce{La_{0.5}Ca_{0.5}MnO3}, ziz-zag spin chains of the CE type magnetic structure arise as a result of the charge and orbital order associated with a checkerboard arrangement of B-site Mn$^{3+}$ and Mn$^{4+}$ ions \cite{1997Radaelli}. Despite considerable departures from the ideal 180{\degree} Mn-O-Mn bonding geometry of up to 40{\degree} (due to octahedral tilts), these magnetic structures can be well understood in terms of dominant B-B magnetic exchange interactions described by the Goodenough-Kanamori-Anderson (GKA) rules \cite{1955Goodenough}. Rare-earth magnetism can play an important role via A-A and A-B interactions, but typically only at low temperatures \cite{2001Munoz_orth} owing to the much weaker exchange between $f-f$ and $f-d$ electrons, respectively, compared to that of $d-d$ electrons.

In the \ce{AMn7O12} quadruple perovskite manganites a large $a^+a^+a^+$ octahedral tilting pattern (in Glazer notation \citep{1972GlazerNotation}) introduces an ordered arrangement of Mn ions onto the A sites of the perovskite framework such that these systems also incorporate A-A and A-B $d-d$ exchange pathways. Compared to the simple perovskites, competition between B-B exchange and these additional interactions can lead to new and complex paradigms in frustrated magnetism, especially because the B-B exchange is diminished due to the large octahedral tilting pattern (Mn-O-Mn bond angles reduced by $\sim$40-45{\degree} away from the ideal 180{\degree} \cite{2016Belik, 2018Zhang_RMn7O12}). For example, when R = La, Ce, Nd, Sm and Eu a collinear ferrimagnetic structure was observed that cannot be explained by dominant B-B interactions alone \citep{2018Johnson_RMn7O12, 2019Johnson_YMn7O12}. For A = Ca, Sr, Cd, Pb and \ce{Na_{1-x}Ca_xMn7O12} more complex magnetic structures were observed; a constant moment magnetic helix with a modulated spin helicity, and an incommensurate pseudo CE-type phase, respectively \cite{2016Johnson, 2017Johnson, 2018Johnson_NaCaMn7O12}. Both incommensurate structures arise from a balance between competing A-A, A-B, and B-B exchange interactions.

Recently, a family of \ce{A2A$'$A$''$B4O12} A-site columnar ordered quadruple perovskite manganites have been synthesised \cite{2017Zhang}. In these manganese oxides a large $a^+a^+c^-$ octahedral tilting pattern gives rise to three crystallographically distinct A sites, and creates a unique set of exchange pathways not found in either the simple or quadruple perovskite manganites described above. Here, both A-B and B-B $d-d$ super-exchange interactions are present, while the $a^+a^+c^-$ tilts remove A-A $d-d$ super-exchange pathways, which are then reduced to super-super-exchange. The relative strength of A-A $d-d$ exchange has recently been demonstrated in \ce{(NaDy)MnMnTi4O12} \cite{2020Liu}, where in the absence of A-B and B-B interactions (B = Ti$^{4+}$) antiferromagnetic order develops on the A-site sublattices only below $\sim$ 12 K. To the contrary, in \ce{Tm2MnMnMn4O12} ferrimagnetic order appears below 74 K, with a ferromagnetic B-site sublattice that is in direct contradiction to the C-type antiferromagnetic B-site sublattice theoretically predicted by the GKA rules --- a state that is instead thought to arise as a result of A-B exchange dominating over both A-A and B-B exchange \cite{2019Vibhakar}. Indeed, it was shown that when A-B exchange is weakened by substituting Cu$^{2+}$ for Mn$^{3+}$ on the A$'$ sites (\ce{R2CuMnMn4O12}, R = Dy and Y) antiferromagnetic spin canting is introduced onto the B-site sublattice \cite{2020Vibhakar}. Hence, it has become clear that the A-site columnar ordered quadruple perovskite manganites present a flexible framework in which novel frustrated geometries of A-A, A-B, and B-B exchange interactions may be tuned through chemical substitution to give rise to unconventional magnetic states.

In this paper, we report the tuning of magnetic exchange interactions in the \ce{Sm2MnMnMn_{4-x}Ti_xO_{12}} A-site columnar ordered quadruple perovskite manganites by chemical substitution of non-magnetic Ti$^{4+}$ for B-site Mn$^{3+}$ for $x=1$, $x=2$ and $x=3$. This series interpolates between the A-site only, antiferromagnetic structure of \ce{(NaDy)MnMnTi4O12} \cite{2020Liu}, and the full ferrimagnetic structure of \ce{Tm2MnMnMn4O12} \cite{2019Vibhakar}, discussed above. \ce{Sm2MnMnMn_{4-x}Ti_xO12} has a tetragonal crystal structure with space group $P4_2/nmc$, where the charge and orbital order responsible for the lower orthorhombic symmetry of the \ce{R2MnMnMn4O12} manganites \cite{2017Zhang, 2019Vibhakar} has been suppressed by the B-site chemical disorder \cite{2019Belik}. The three symmetry inequivalent A sites, labelled $A$ (Sm1), $A'$ (Mn1) and $A''$ (Mn2), are occupied by 10-fold coordinated Sm$^{3+}$, square planar coordinated Mn$^{2+}$ or Mn$^{3+}$, depending on the amount of Ti$^{4+}$ substituted on the B sites, and by tetrahedrally coordinated Mn$^{2+}$ respectively. The B sites, labelled Mn3/Ti1, are octahedrally coordinated and occupied by 75\% Mn$^{3+}$ and 25\% Ti$^{4+}$ for the $x = 1$ sample, 50\% Mn$^{3+}$ and 50\% Ti$^{4+}$ for the $x = 2$ sample and 25\% Mn$^{2+}$ and 75\% Ti$^{4+}$ for the $x = 3$ sample. The two cations are disordered across the B sites with an average oxidation state of between +3.25 and +3.5. Published DC magnetometry measurements show that samples with $x = 1$, $x = 2$ and $x = 3$ undergo a single magnetic phase transition at temperatures of $T_c$ = 27 K, 62 K and 40 K respectively \cite{2019Belik}. Here we perform a quantitative neutron diffraction study and show that all compositions undergo a transition to a long range ordered magnetic phase below $T_c$ in which the Mn ions adopt a collinear ferrimagnetic structure. Remarkably these results demonstrate an unusual increase in the ordering temperature between the $x =1$ and $x = 2$ samples, despite an increase in the amount of non-magnetic Ti$^{4+}$ on the B-site sublattice -- contrary to what is observed in numerous other magnetically dilute systems \cite{1962Elliott} and is suggestive of a departure from mean field physics. In the $x=2$ sample we demonstrate that site-selective magnetic dilution has enabled the tuning of ferrimagnetic compensation and the introduction of temperature-induced magnetization reversal. We show that in the $x = 3$ sample, long-range B-site magnetic order must percolate via \emph{both} A-B and B-B exchange, demonstrating the importance of A-B exchange in mediating magnetism in the A-site columnar ordered quadruple perovskites.

The paper is organised as follows. In Sec. \ref{SEC::experimentaldetails} we present the experimental details, followed by the results of the neutron powder diffraction refinement, crystal electric field and magnetic anisotropy calculations and percolation calculations in Sec. \ref{SEC::npd}, \ref{SEC::CryElecField} and \ref{SEC::percolation} respectively. A discussion on the variation of ordering temperature and size of the magnetic moments with Ti$^{4+}$ content is given in Sec. \ref{SEC::discussion} and a summary of our findings in Sec. \ref{SEC::conclusions}.

\section{Experimental Details}\label{SEC::experimentaldetails}
Polycrystalline samples of \ce{Sm2MnMnMn_{4-x}Ti_xO12} were synthesised under high-temperature and high-pressure conditions from stoichiometric mixtures of \ce{Mn2O3}, \ce{Mn3O4}, \ce{TiO2} and \ce{Sm2O3} for $x = 1$, from \ce{Mn2O3}, \ce{MnO}, \ce{TiO2} and \ce{Sm2O3} for $x = 2$, and \ce{MnO}, \ce{TiO2} and \ce{Sm2O3} for $x = 3$, as detailed in Ref. \citenum{2019Belik}. Neutron powder diffraction experiments were performed using the time of flight diffractometer, WISH at ISIS \cite{2011ChaponWISH}. $x = 1$, $x= 2$ and $x = 3$ samples of mass 0.8g, 0.9g, and 0.7g, respectively, were loaded into 3mm vanadium cans and cooled to 1.5 K. Diffraction data with good counting statistics were collected on warming from a base temperature of 1.5 K up to 35 K in 5 K steps for the $x = 1$ sample, up to 78 K in 6 K steps for the $x = 2$ sample, and up to 60 K in 6 K steps for the $x = 3$ sample. Measurements with high counting statistics were made in the paramagnetic phase (80 K for $x = 1$ and $x = 3$, 90 K for $x = 2$) and at 1.5 K for all three samples. Symmetry analysis was performed using the \textsc{isotropy} software suite \cite{2006Isodistort}, and the diffraction data were fit using \textsc{fullprof} \cite{1993Rodriguez}. Magnetic measurements were performed on a SQUID magnetometer (Quantum Design, MPMS-XL-7T) between 2 and 400 K in 100 Oe under both zero-field-cooled (ZFC) and field-cooled on cooling (FCC) conditions. Isothermal magnetization measurements were performed between $-70$ and 70 kOe at 5 K. 

\begin{figure*}[ht]
\centering
\includegraphics[width=\linewidth]{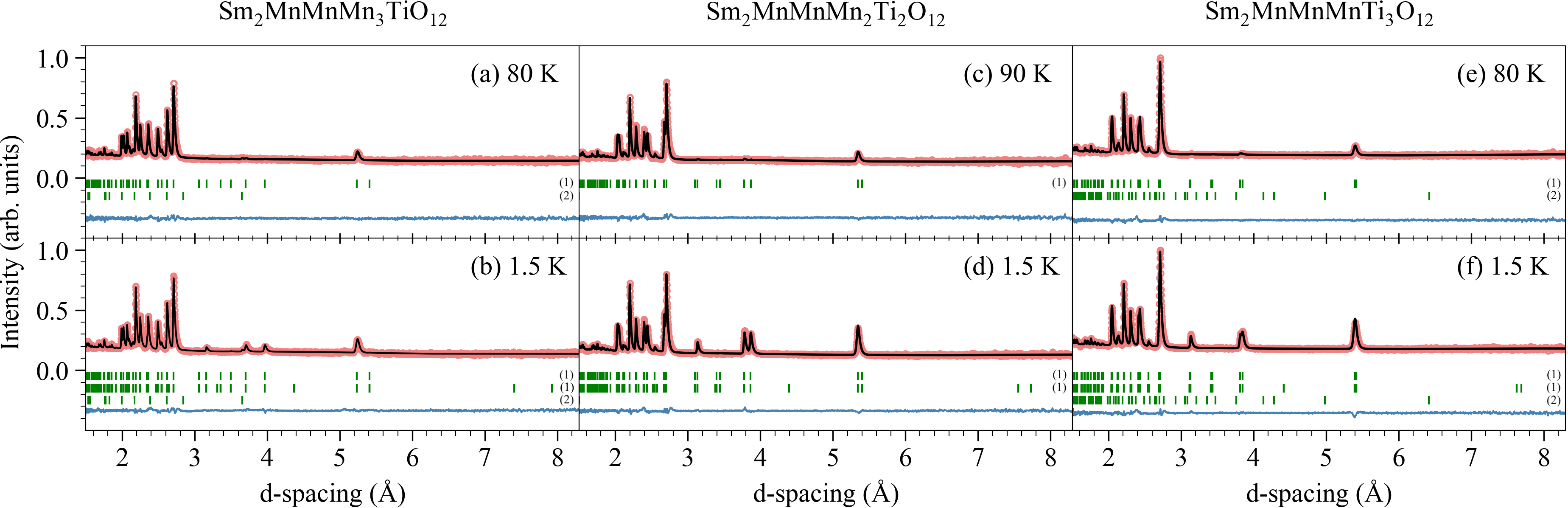}
\caption{\label{neutron}Neutron powder diffraction data collected on the time of flight diffractometer WISH at ISIS of (a)-(b) the x = 1 sample (c)-(d) the x = 2 sample and (e)-(f) the x = 3 sample, at temperatures representative of the paramagnetic phase and at 1.5 K. The data are given by the red circles, the fit of the data by the solid black line, and their difference as a blue line. The green tick marks are the nuclear and magnetic reflections from top to bottom of \ce{Sm2MnMnMn_{4-x}Ti_xO_{12}} labelled (1) and an impurity phase labelled (2).}
\end{figure*}

\section{Results}\label{SEC::results}
\subsection{Crystal and Magnetic Structures}\label{SEC::npd}

The $P4_2/nmc$ crystal structure model previously reported for \ce{Sm2MnMnMn_{4-x}Ti_xO12} \cite{2019Belik} was refined against the neutron powder diffraction data measured in the paramagnetic phase of each of the samples. For the x = 1 and x = 3 samples weak diffraction peaks that could not be indexed by the $P4_2/nmc$ model were found to originate in an \ce{MnCO3} impurity phase (0.11 wt. \%) and \ce{Sm2Ti2O7} impurity phase (0.25 wt\%) respectively. For all three samples the thermal parameter of Mn1 refined to unphysically large values when the ion was constrained to the centre of the square planar coordination (Wyckoff position 2a). Hence, we adopted a split-atom model representative of site disorder above and below the square planar coordination (Wyckoff position 4c), as employed in Ref. \citenum{2019Belik}, which reduced the thermal parameter to physical values. Sm was free to substitute Mn on the $A'$ and $A''$ sites and Mn was free to substitute Sm on the $A$ sites, while imposing full site occupations. An excellent fit was achieved in all samples and the refined stoichiometries were as follows: \ce{Sm_2Mn(Mn_{0.84}Sm_{0.16})Mn3TiO12} ($R = 2.72\%$, $wR = 2.90\%$, $R_\mathrm{Bragg} = 4.06\%$ at 80 K), \ce{Sm_2Mn(Mn_{0.93}Sm_{0.07})Mn2Ti2O12} ($R = 2.71\%$, $wR = 2.87\%$, $R_\mathrm{Bragg} = 5.05\%$ at 90 K) and \ce{Sm_2MnMnMnTi3O12} ($R = 2.03\%$, $wR = 2.20\%$, $R_\mathrm{Bragg} = 5.69\%$ at 80 K). Note that the $R_\mathrm{Bragg}$ values are large in all samples due to strong absorption from Sm$^{3+}$ which results in much weaker diffracted intensities.

\begin{figure}[ht]
\includegraphics[width=\linewidth]{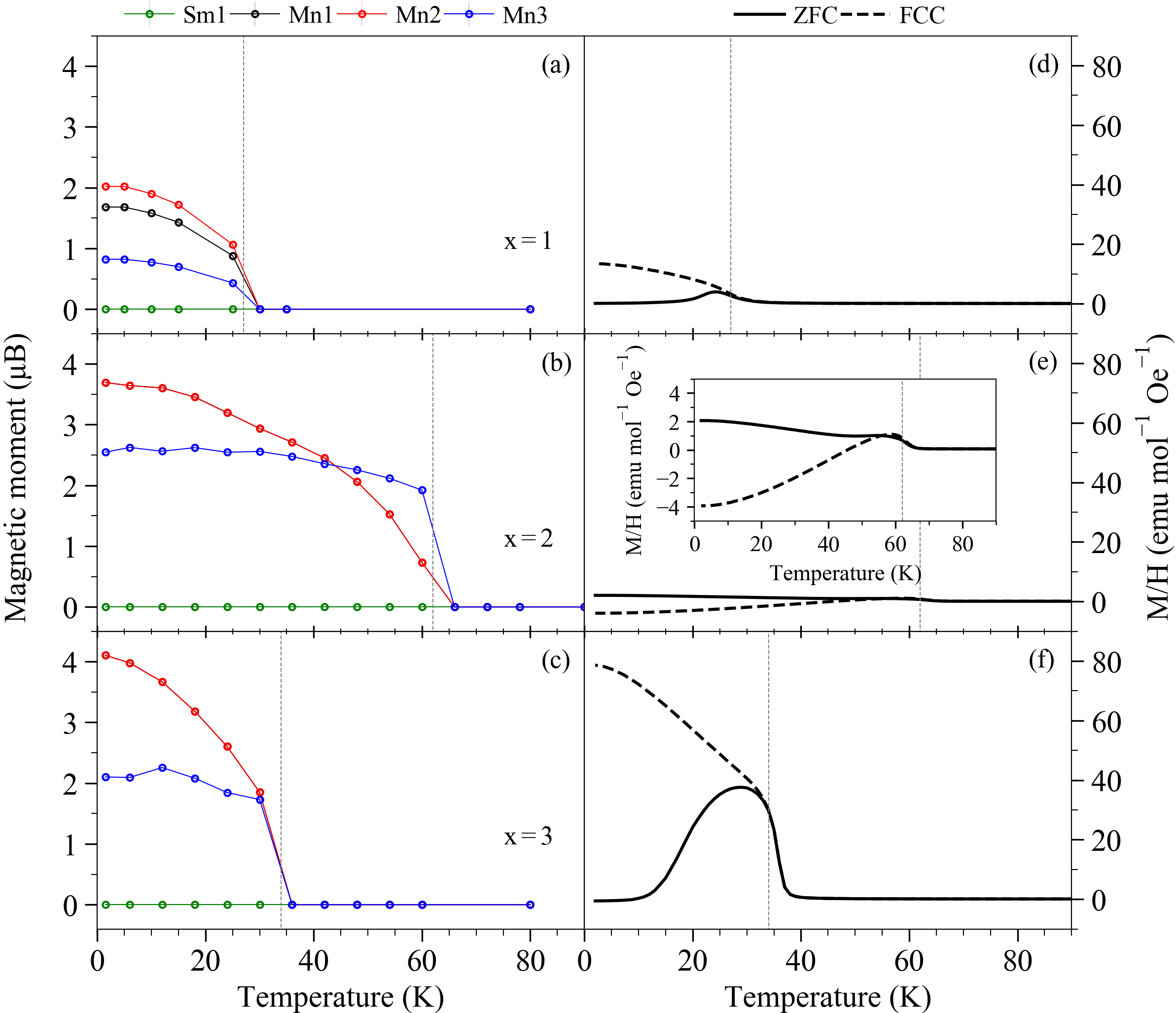}
\caption{\label{FIG::tempdep} (a)-(c) The temperature dependence of ionic moment magnitudes given for each of the symmetry inequivalent magnetic ions. The moments of Mn1 and Mn2 were fixed to be the same in the $x=2$ and $x=3$ samples to be consistent with their equivalent nominal oxidation states. (d)-(f) ZFC and FCC magnetization measurements under an applied DC field of 100 Oe.  The inset in (e) shows the ZFC and FCC measurements for the $x= 2$ sample on a reduced scale.}
\end{figure}

A gradual growth of the (200), (002), (112) and (110) reflections was observed for all three samples below $T_c$ in accordance with changes in the sample's magnetization (Fig. \ref{FIG::tempdep}), and hence these peaks were identified as magnetic in origin. For all three samples, the magnetic peaks are sharp and of a similar width to the nuclear peaks, indicating that the magnetic structure is well correlated. The above peak positions were consistent with a $\Gamma$-point, $\textbf{k} = (0,0,0)$, magnetic propagation vector. The magnetic $\Gamma$-point representation for the Wyckoff positions of the Sm and Mn sublattices decomposes into six 1-dimensional irreducible representations (irreps) ($\Gamma_1^\pm$, $\Gamma_2^\pm$, $\Gamma_3^+$ and $\Gamma_4^+$) and two 2-dimensional (2D) irreps ($\Gamma_5^\pm$), which together define a total of 12 different magnetic symmetries. The rapid increase of the magnetic susceptibility below $T_c$ (Fig. \ref{FIG::tempdep}) indicated the presence of ferromagnetic sublattices, which are only consistent with the $\Gamma_3^+$ or $\Gamma_5^+$ irreps. Furthermore, the observation of finite magnetic intensity on the (002) reflection indicates that the magnetic moments are not aligned parallel to the c-axis (as magnetic diffraction intensity is proportional to the magnetic moment components perpendicular to the scattering vector), which uniquely identifies that the magnetic structure transforms by the $\Gamma_5^+$ irrep. 

\begin{figure}[ht]
\includegraphics[width=\linewidth]{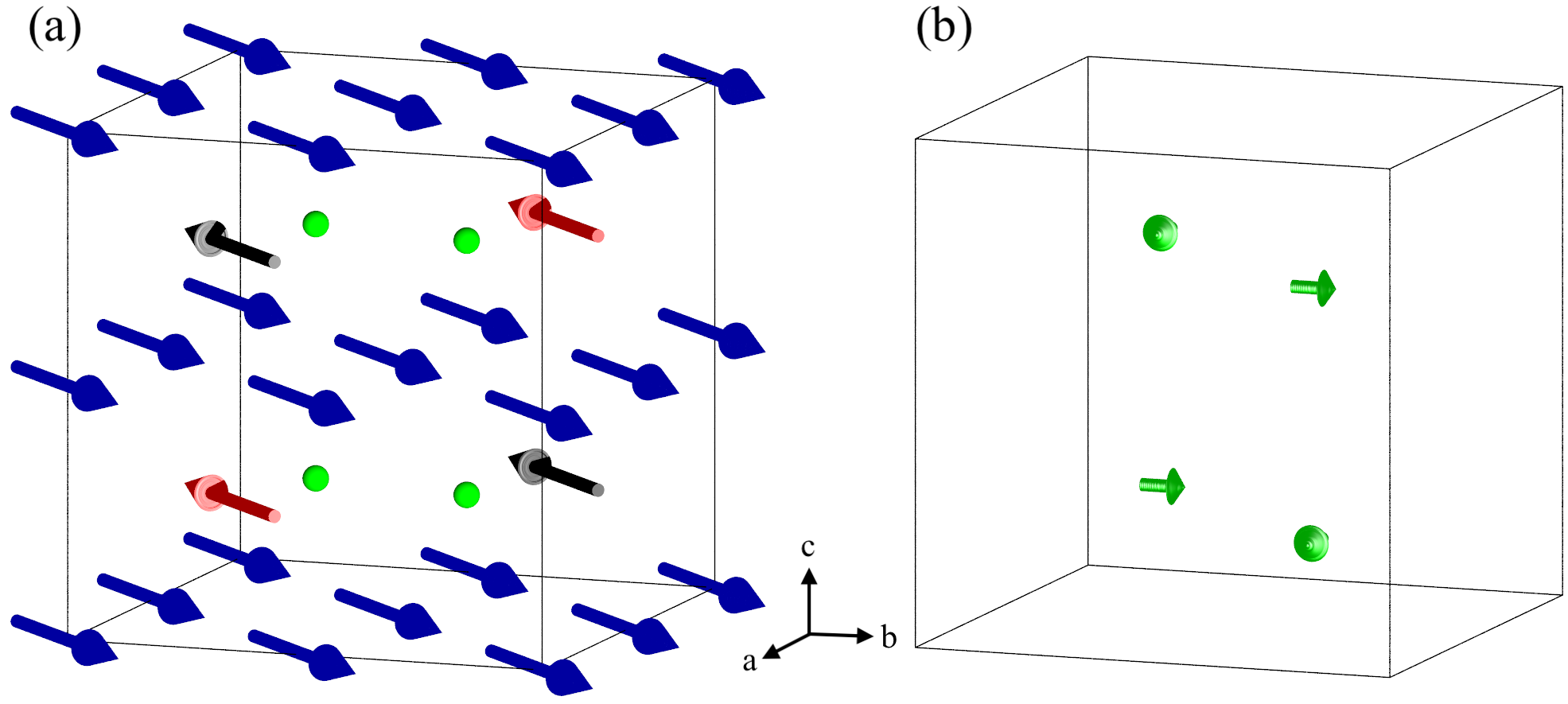}
\caption{\label{FIG::magstruc}(a) Experimentally determined magnetic structure of \ce{Sm2MnMnMn_{4-x}Ti_xO_{12}} upto a direction in the $ab$ plane. (b) Theoretically proposed spin configuration of the Sm sites as determined by point charge calculations. The $A$ sites are colored in green, the $A'$ sites in black, and the $A''$ sites in red. The B sites are colored in blue, and for clarity we show the scenario in which all B-site ions of a magnetic unit cell are Mn3 -- a more accurate depiction of the unit cell would consist of statistically disordered Ti1 and Mn3 ions on the B-sites. The magnetic space group is Ccc$'$a$'$ (No. 68.516) with basis vectors (1,1,0),(-1,1,0),(0,0,1) and an origin of (1,1/2,1) with respect to the paramagnetic space group. }
\end{figure}

\begin{figure}[ht]
\includegraphics[width=\linewidth]{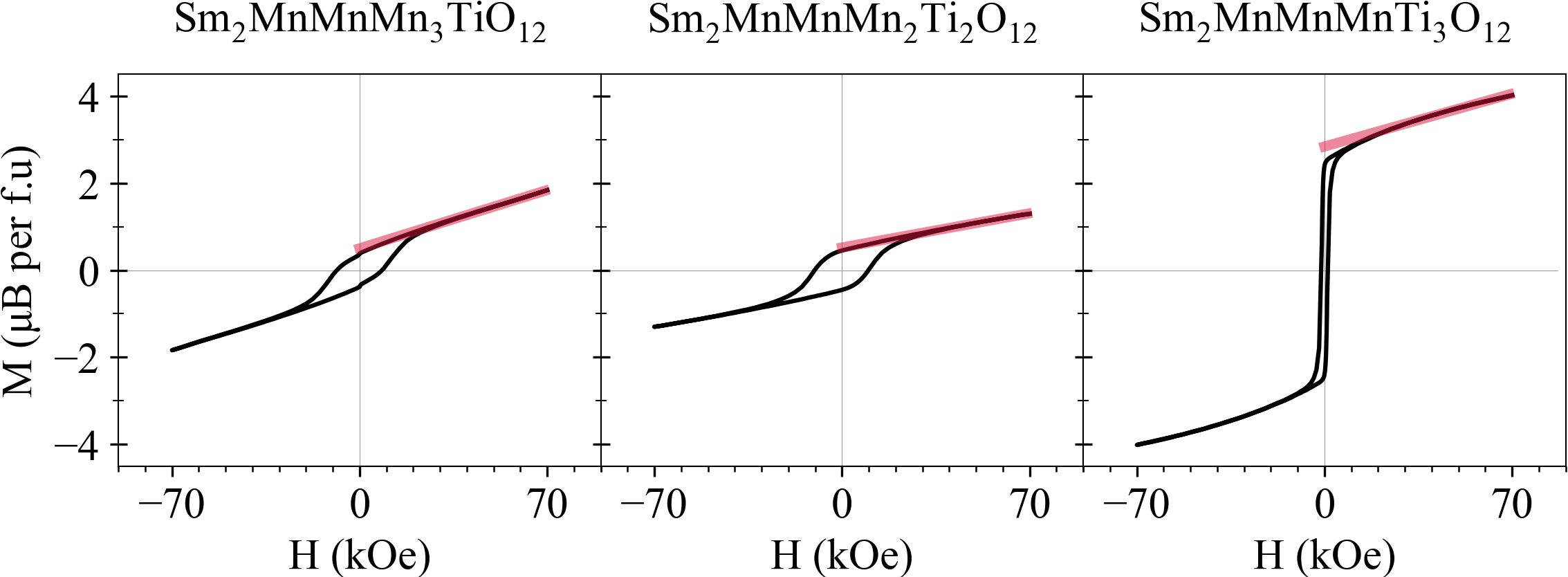}
\caption{\label{FIG::MHcurves} Plots of the magnetization, M, as a function of applied field, H, taken at a temperature of 5 K for each of the samples. A magnetic field range of 22.5 to 70 kOe was used to extrapolate the magnetization to zero field and the extrapolated remnant magnetization is shown by the red line.}
\end{figure}

The $\Gamma_5^+$ symmetry adapted basis functions for the $A$, $A'$, $A''$ and B sites are summarised in Tables \ref{TAB::Asites}, \ref{TAB::A'A''sites}, and \ref{TAB::Bsites}. Magnetic structure models constructed from linear combinations of these basis functions were exhaustively tested by refinement against the neutron diffraction data measured at 1.5 K from all samples. A model in which the B-sites moments were oriented collinearly in the $ab$ plane and ferromagnetically coupled, and the $A'$ and $A''$ sites collinear but anti-aligned with respect to the B site sublattice, gave the best fit to the data ($R_{Mag}$ =  17.0\% for x = 1, 5.39\% for $x = 2$ and 4.34\% for x = 3), and is shown in Fig. \ref{FIG::magstruc}a. The refined moment magnitudes are given in Table \ref{TAB::magmoments}, which also lists the net ferrimagnetic moment per formula unit. The strong dependence of the net ferrimagnetic moment on composition, $x$, is reflected in the remnant magnetisation, $M_r$, evaluated by extrapolating the powder-averaged bulk magnetisation to zero field (see Figure \ref{FIG::MHcurves}). Scaled extrapolated values are listed in Table \ref{TAB::magmoments}, and their relative magnitudes correspond well to the net ferrimagnetic moment per formula unit obtained from neutron diffraction, except for the $x = 2$ sample. This discrepancy cannot be explained by the proposed magnetic structure model.

The refined magnetic structure is consistent with the collinear ferrimagnetic phase found in \ce{Tm2MnMnMn4O12} \cite{2019Vibhakar} and the FI phase of \ce{R2CuMnMn4O12} \cite{2020Vibhakar}. Moment orientations of $m||[x,0,0]$, $m||[x,x,0]$ or $m||[x,y,0]$ correspond to different magnetic symmetries described by the three distinct order parameter directions of the $\Gamma_5^+$ irrep (see Tables \ref{TAB::Asites}, \ref{TAB::A'A''sites} and \ref{TAB::Bsites}), however these three cases cannot be differentiated by our neutron diffraction data due to powder averaging in the tetragonal crystal symmetry. In the following we adopt a model with $m||[x,x,0]$, which is consistent with Sm crystal electric field calculations (Sec. \ref{SEC::CryElecField}).

The temperature dependences of all symmetry-inequivalent moment magnitudes, plotted in Fig. \ref{FIG::tempdep}, were evaluated by fitting the magnetic structure model to variable temperature neutron powder diffraction data measured from all three samples. The refinements were constrained such that each magnetic ion had a moment no larger than its theoretically predicted maximum, given in Table \ref{TAB::magmoments}, and with the ratio of the Mn1:Mn2 moments chosen to be consistent with their nominal oxidation states. The refinement was insensitive to a Sm1 moment at all temperatures, so it was set to zero. For the $x = 2$ sample the B-site moments grow rapidly as the temperature decreases, but quickly saturate in comparison to the the A-site moments which steadily grow with decreasing temperature. As a result, the net magnetisation (the sum over all magnetic moments in the unit cell with the A sublattices antialigned with respect to the B sublattice) reverses direction with decreasing temperature, giving rise to a compensation point at 45$\pm$3 K. Magnetisation reversal was also observed in the temperature dependent magnetic susceptibility FCC data of the $x=2$ sample, as shown in the inset of Fig \ref{FIG::tempdep}e, for which the compensation temperature was determined to be 46$\pm$ 1 K in excellent agreement with the neutron diffraction results. We note that the absolute value of the negative, low temperature magnetic susceptibility is not reflective of the full net ferrimagnetic moment, as the coercive field of the sample may come close to the 100 Oe applied field.

\begin{table}[ht]
\caption{\label{TAB::magmoments}Absolute values of the 1.5 K experimental (Exp.) and theoretical (Calc.) ionic magnetic moments of \ce{Sm2MnMnMn_{4-x}Ti_xO_{12}}. The net magnetisation, M, per formula unit is determined from the experimental moment values, where the A sites align antiparallel to the B sites. The remnant magnetization, M$_r$, extrapolated from magnetometry data was scaled by a factor of 2 to account for powder averaging. All moments are given in units of $\mu$B.}
\centering
\begin{ruledtabular}
\begin{tabular}{ c | c | c | c | c | c | c }
Ion & \multicolumn{2}{c |}{x = 1}  &  \multicolumn{2}{c |}{x = 2} & \multicolumn{2}{c}{x = 3}  \\
 & {Exp.} & Calc. & {Exp.} & {Calc.} & {Exp.} & {Calc.} \\
\hline
Sm1 &  0.00     & 0.71   &  0.00    & 0.71   &  0.00    & 0.71   \\
Mn1 &  1.70(7)  & 4.0    &  3.72(7) & 5.0    &  4.09(8) & 5.0    \\
Mn2 & 2.03(9)   & 5.0    &  3.72(7) & 5.0    &  4.09(8) & 5.0    \\
Mn3 & 0.82(6)   & 4.0    &  2.54(6) & 4.0    &  2.1(1)  & 5.0    \\
\hline
M per f.u & \multicolumn{2}{c|}{1.3(1)}  & \multicolumn{2}{c|}{2.4(1)} & \multicolumn{2}{c}{6.0(2)} \\            
\hline
M$_r$ per f.u & \multicolumn{2}{c|}{1.006(7)} & \multicolumn{2}{c|}{1.05(1)} & \multicolumn{2}{c}{5.68(3)}\\
\hline
$T_c$ &  \multicolumn{2}{c|}{27 K} & \multicolumn{2}{c|}{62 K} & \multicolumn{2}{c}{34 K} \\
\end{tabular}
\end{ruledtabular}
\end{table}

\subsection{Magnetic Anisotropy}\label{SEC::CryElecField}
The Jahn-Teller active Mn$^{3+}$ ions present in the $x = 1$ and $x = 2$ samples are expected to impose an average weak single-ion anisotropy $||c$ \cite{2015Whangbo, 2019Vibhakar}, while all other Mn ions are isotropic Mn$^{2+}$. The three \ce{Sm2MnMnMn_{4-x}Ti_xO_{12}} samples have a magnetization direction in the $ab$ plane, perpendicular to the easy axis of the Mn$^{3+}$ ions, suggesting that the magnetic anisotropy of Sm$^{3+}$ may instead play an important role despite our data being insensitive to an ordered moment on the Sm sites (the predicted Sm$^{3+}$ moment of $\sim$ 0.7$\mu$B is feasibly below the sensitivity of our neutron diffraction experiment). We note that in \ce{Tm2MnMnMn4O12} the competition between the Mn$^{3+}$ and RE magnetic anisotropies was found to result in a spin reorientation phase transition. If the magnetic anisotropy is indeed determined by Sm$^{3+}$, then the absence of a spin-reorientation phase transition in the \ce{Sm2MnMnMn_{4-x}Ti_xO_{12}} compounds implies significant $f-d$ exchange coupling immediately below $T_c$ (N.B. in the orthoferrites strong $f-d$ exchange coupling is also present in \ce{SmFeO3} and leads to a phase transition approx. 300 K - 400 K above all other rare-earth orthoferrites \cite{1969White}).

Point charge approximations to the Sm$^{3+}$ crystal electric field (CEF) were used to calculate the ground state magnetic anisotropy of Sm$^{3+}$, and hence establish whether $f-d$ exchange coupling of rare-earth moments to the manganese sublattices might be responsible for the $ab$ plane magnetization direction observed in \ce{Sm2MnMnMn_{4-x}Ti_xO_{12}}. Sm$^{3+}$ has electronic configuration [Xe]4$f^{5}$, and Hunds rules give S = 5/2, L = 5 and J = 5/2 for the lowest-energy 6-fold multiplet of states. The local crystal electric field of Sm$^{3+}$ has the point group symmetry $2mm$, which lifts the multiplet degeneracy to form 3 Kramers doublets. All non-zero crystal electric field parameters, $B_n^m$ \cite{1964Hutchings}, were calculated for Sm$^{3+}$ in its two ionic positions, ($\frac{1}{4}$, $\frac{1}{4}$, $z$) and ($\frac{1}{4}$, $\frac{1}{4}$, $z + \frac{1}{2}$), labelled Sm11 and Sm12 respectively (Table \ref{TAB::CEF parameters}). Ionic positions refined for the $x = 2$ compound at 1.5 K were used for reference, and similar values were obtained using the ionic positions of the $x = 1$ and $x = 3$ samples. Sm11 and Sm12 are symmetry equivalent, and related by a $4_2$ screw, therefore all values of $B_n^m$ are the same across the two sites, except the $B_2^2$ and $B_4^2$ terms which change sign upon 90$^\circ$ rotation about $c$. 

We consider the following Hamiltonian for a single Sm$^{3+}$ ion;

\begin{equation}
\mathcal{H} = \sum_{n}\sum_{m=-n}^{n}B_n^m O_n^m + g_J\mu_\mathrm{B}\mathbf{J}\cdot\mathbf{B}_\mathrm{ex}
\label{EQN::singleionhamiltonian}
\end{equation}
where $O_n^m$ are the Stevens operators \cite{1964Hutchings}, $g_J$ is the Land\'{e} $g$-factor, $\mathbf{J} = (J_x, J_y, J_z)$ is the vector total angular momentum operator, and $\mathbf{B}_\mathrm{ex}$ is an effective exchange field originating in $f-d$ exchange interactions with the manganese sublattices.

The magnetic anisotropy of Sm11 and Sm12 was determined by repeatedly solving Equation \ref{EQN::singleionhamiltonian} for a 1 T field applied in different directions over a full hemisphere. Eigenfunctions were found through diagonalisation of $\mathcal{H}$, and magnetic moment components $m_x$, $m_y$, and $m_z$ were calculated by summing the expectation values of $\langle J_x\rangle$, $\langle J_y\rangle$ and $\langle J_z\rangle$ over all eigenfunctions, multiplied by $g_J$, and weighted by Boltzmann statistics evaluated at T = 10 K.

\begin{figure}[ht]
\includegraphics[width=\linewidth]{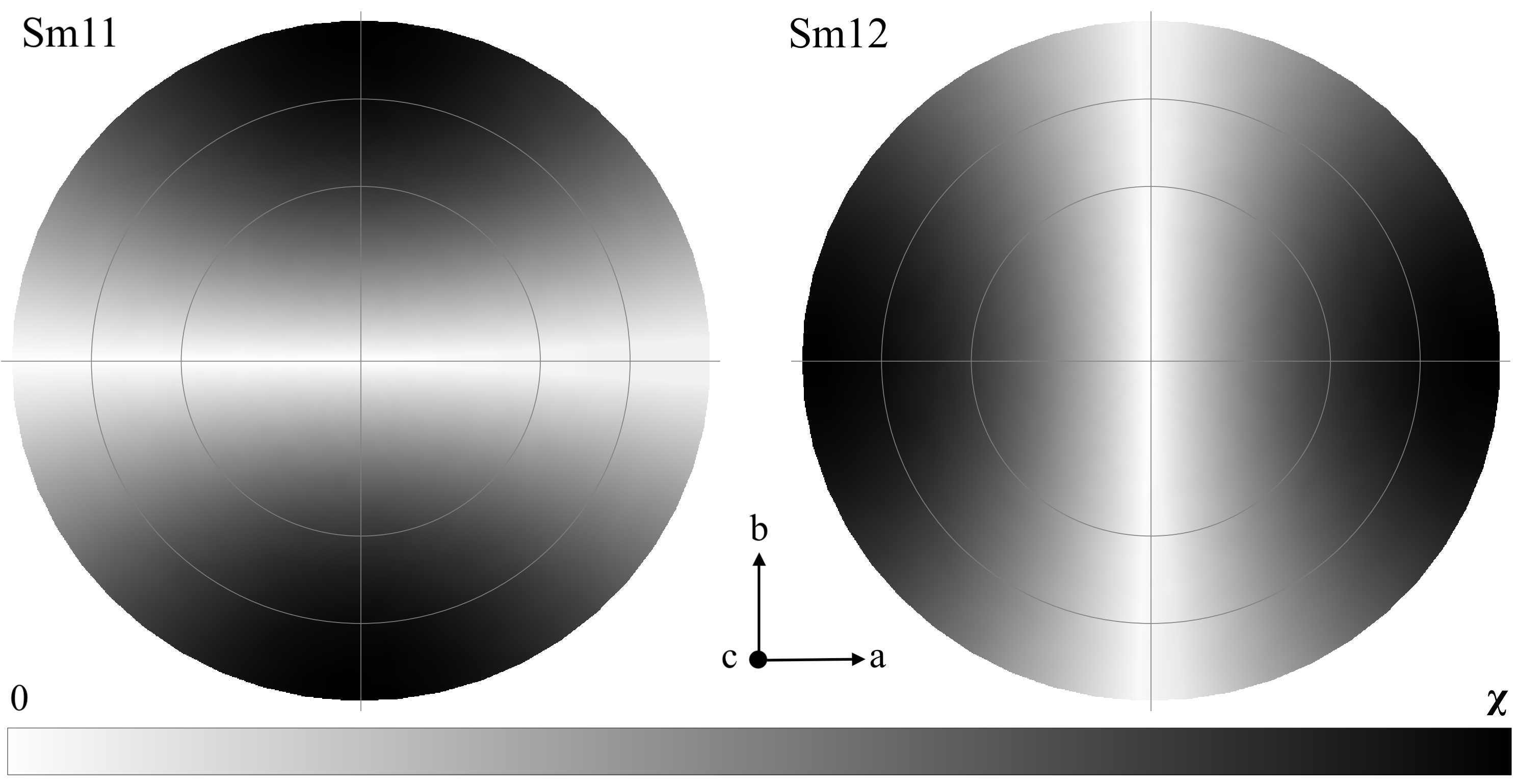}
\caption{\label{FIG::pointchargecalcs} The magnetic susceptibility of Sm11 and Sm12 plotted as a stereographic projection over a full hemisphere under an applied field of 1T.}
\end{figure}

The total Sm11 and Sm12 moments for the different field directions are shown in Fig. \ref{FIG::pointchargecalcs} as a stereographic projection. A strong Ising like anisotropy was found $||y$ for Sm11 and $||x$ for Sm12. The easy axis rotates between x and y for Sm$^{3+}$ sites in accordance with the $4_2$ screw symmetry that relates the two sites. We note that this perpendicular alignment of Sm11 and Sm12 moments parallel to $y$ and $x$, respectively, is described by the B1 symmetry adapted mode associated with the (a,a) order parameter direction of $\Gamma_5^+$ (see Table \ref{TAB::Asites}). Within this symmetry, the Sm moments only couple to the Mn collinear ferrimagnetic structure if the Mn moments lie $||[x,x,0]$. Hence, not only can the Sm ions impose an $ab$-plane magnetic anisotropy, they can also determine the direction of the magnetization in the plane (as shown in Fig. \ref{FIG::magstruc}b).

\begin{table}[ht]
\caption{\label{TAB::CEF parameters}
Non-zero crystal electric field parameters for Sm$^{3+}$ in \ce{Sm2MnMnMn2Ti2O12} evaluated by the point charge model, and given in units $\mu$eV.}
\begin{ruledtabular}
\begin{tabular}{c | c | c c c c c }
Ion & Frac. coords. & $B_2^0$ & $B_2^2$ & $B_4^0$ & $B_4^2$ & $B_4^4$\\
\hline
Sm11 & 0.25, 0.25, 0.209  &   2430 & 5204 & -11.65 & 40.94 & 41.85 \\
Sm12 & 0.25, 0.25, 0.709  &   2430 &   -5204 & -11.65 &  -40.94  & 41.85  \\
\end{tabular}
\end{ruledtabular}
\end{table}

\subsection{Percolation Calculations}\label{SEC::percolation}

First, we consider the percolation of long-range magnetic order via B-B exchange alone. In \ce{Sm2MnMnMn_{4-x}Ti_xO12} B-B exchange pathways span a three dimensional framework. Each B site has six nearest-neighbour B sites, such that for every non magnetic Ti$^{4+}$ion substituted onto this sublattice, six nearest neighbour magnetic B-B exchange pathways are removed. The site-percolation threshold (the probablility, p, of Mn B site occupation required to establish a percolating cluster that spans the entire lattice between two opposing faces) for a simple cubic lattice representative of the B-B framework is 31.16\% \cite{1985Stauffer}. Hence, B-B exchange is expected to play no role in establishing long-range magnetic order in the $x=3$ sample, for which the B site Mn occupation is 25\%. 

To explore the origin of long range magnetic order in the $x=3$ sample we calculated the percolation threshold if magnetic order were to percolate via both A-B and B-B exchange. Two interpenetrating simple cubic lattices were used to represent the \ce{A2A$'$A$''$B4O12} unit cell, where the Mn $A'$ and $A''$ sites were set to be fully occupied, and the Sm $A$-sites were set to be unoccupied to exclude interactions with the RE ions. Mn ions were randomly placed on the B sites according to a probability of occupation, p. For a given p, the sites were sorted into clusters, where a cluster is composed of a continuous network of nearest neighbour Mn ions, separated from other clusters by Ti ions. The percolation threshold was evaluated as the minimum value of p for which there exists a cluster that percolates through the entire lattice along the c-axis.

\begin{figure}[ht]
\includegraphics[width=\linewidth]{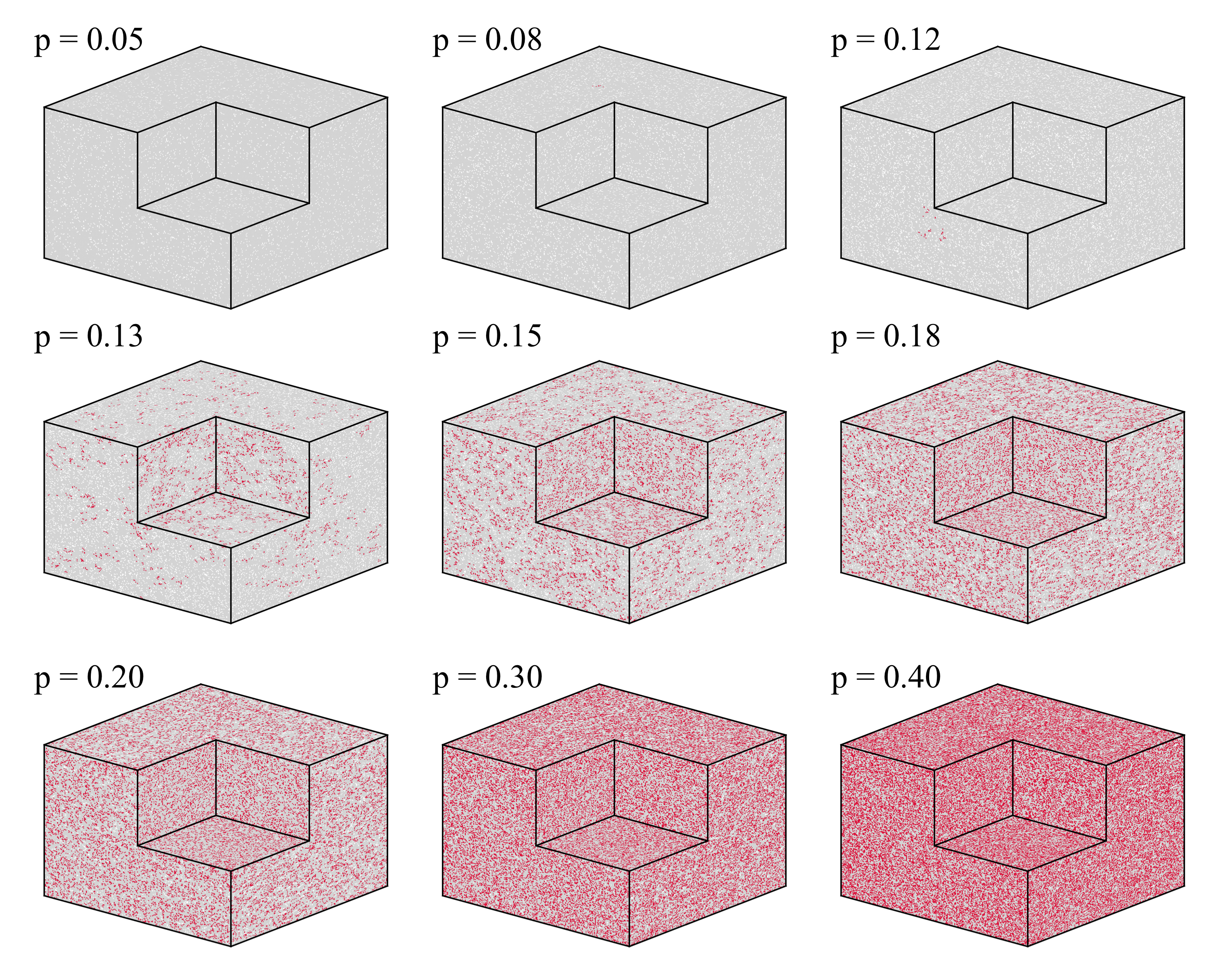}
\caption{\label{FIG::percolation} A three dimensional illustration of a cubic lattice, spanning a hundred \ce{A2A$'$A$''$B4O12} unit cells in each of the three Cartesian directions and showing only clusters of B-site ions when magnetic order is percolated by both A-B and B-B exchange, given for nine different values of the lattice occupation, p. In each figure, only voxels that intersect the surface are shown, and a quarter of the cubic lattice has been removed to reveal the interior. The largest Mn cluster is given in red, all other sites occupied by Mn are given in white and sites occupied by Ti$^{4+}$ are in grey.}
\end{figure}

We found that in this case the percolation threshold was reduced from 31.16\% to 12.4(1)\%, as shown in Fig. \ref{FIG::percolation}, demonstrating that A-B exchange is crucial to mediating long range magnetic order in the $x = 3$ sample.

We note that A-A exchange could also percolate long range magnetic order, however measurements of \ce{(NaDy)Mn2Ti4O12} demonstrated that it does so at temperatures of $\sim$12 K \cite{2020Liu} which is significantly below the ordering temperatures of all measured \ce{Sm2MnMnMn_{4-x}Ti_xO12} samples.

\section{Discussion}\label{SEC::discussion}
\begin{figure}[ht]
\includegraphics[width=\linewidth]{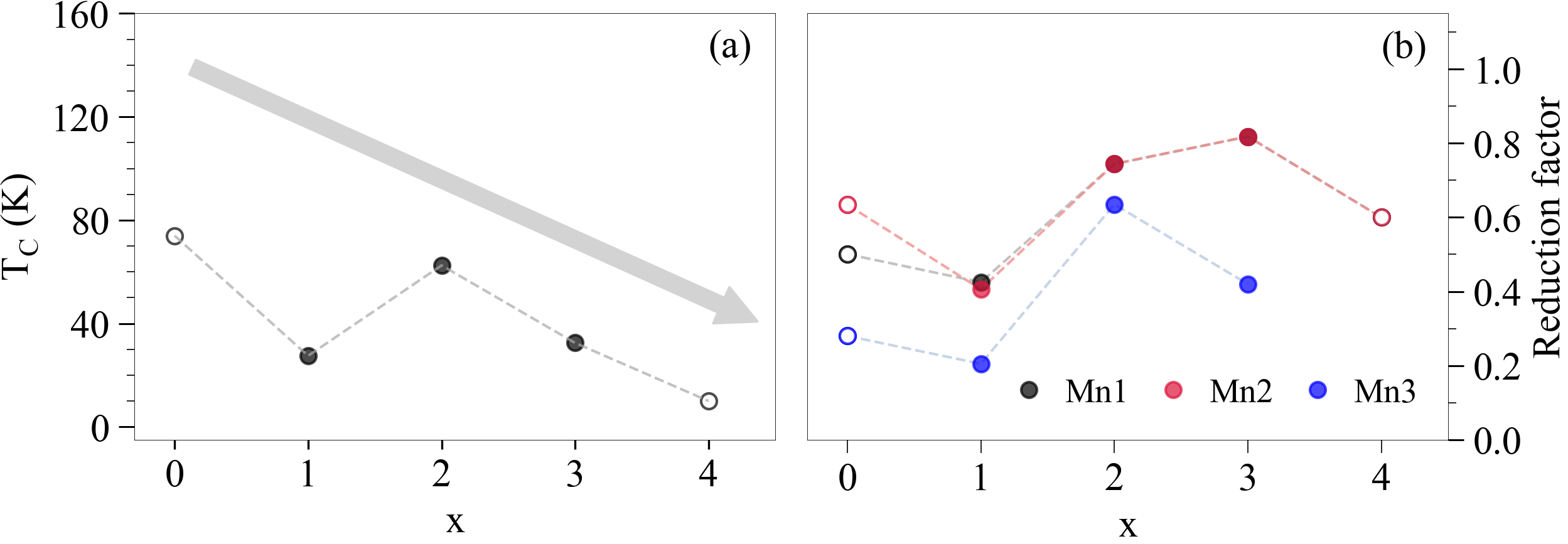}
\caption{\label{FIG::fxn_x} The dependence of (a) the ordering temperature $T_c$ and (b) the reduction in magnetic moment relative to the full $g \langle J_z \rangle \mu_\mathrm{B}$ ground state value, on the concentration of non-magnetic Ti$^{4+}$ (labelled by x). The data for the x = 0 and x = 4 samples are from \ce{Tm2MnMnMn4O12} \cite{2019Vibhakar} and \ce{(NaDy)MnMnTi4O12} \cite{2020Liu}. The grey arrow in (a) is used to illustrate the trend of the mean field dependence of the ordering temperature with x.}
\end{figure}

Diluting a magnetic sublattice with non magnetic ions is predicted to decrease the magnetic ordering temperature \cite{1962Elliott, 1980Landau}, and at the mean field level one expects an approximately linear dependence in the transition temperature on magnetic ion concentration \cite{1975Zvyagin, 1991Cheong, 2014Niven}. Remarkably, the ordering temperature of the \ce{Sm2MnMnMn_{4-x}Ti_xO_{12}} solid solution does not smoothly decrease upon increasing $x$ (Figure \ref{FIG::fxn_x}a), which indicates a significant departure from mean field physics in this system \cite{1945VanVleck} that is also reflected in the moment magnitudes. In insulating materials the magnetic moment of a given ion should reach its full $g \langle J_z \rangle \mu_\mathrm{B}$ value in the ground state. The ground state magnetic moments refined for \ce{Sm2MnMnMn_{4-x}Ti_xO_{12}} are compared to their theoretical values in Table \ref{TAB::magmoments}, and plotted in terms of a relative reduction factor in Figure \ref{FIG::fxn_x}b. A considerable departure from the full moment is observed for all \ce{Sm2MnMnMn_{4-x}Ti_xO_{12}} samples, and the relative reduction in moment magnitude (Figure \ref{FIG::fxn_x}b) correlates well with the non-linear variation observed in $T_\mathrm{c}$ (Figure \ref{FIG::fxn_x}a). Taken together, these results demonstrate the presence of strong spin fluctuations and/or static disorder in the $x=1$ and $x=3$ samples, which are weakened in the $x=2$ sample.

Ideally we would be able to perform the same analysis on a $x = 0$ sample, which has a fully magnetic B-site sublattice, and a $x = 4$ sample, which has a non-magnetic B-site sublattice. However, the former can only be grown in very small quantities \cite{2017Zhang}, and the latter requires the A-site Mn ions to adopt an unfeasible +1 oxidation state. Instead, we have used the moment magnitudes and ordering temperature for \ce{Tm2MnMnMn4O12}, as representative of the $x = 0$ sample, and the ordering temperature and moment magnitude of \ce{(NaDy)MnMnTi4O_{12}} as representative of the $x = 4$ sample, plotted in Fig. \ref{FIG::fxn_x} as unfilled circles. Importantly, these data further support the correlation between the reduction in the relative moment magnitudes and the ordering temperature - an effect that appears to be prevalent in the broader family of materials.

In \ce{Sm2MnMnMn_{4-x}Ti_xO_{12}} static disorder of the magnetic structure may originate in at least one of two forms of crystallographic disorder inherent to these compounds; i) mixed cation occupation and ii) displacements of Mn1 ions above and below their coordination plane. All nuclear neutron diffraction peaks were found to be sharp and limited by the instrumental resolution, indicating that either form of crystallographic disorder statistically effected every unit cell equally, giving a crystal structure that is well correlated on average. The magnetic diffraction peaks had widths similar to those of the nuclear peaks, indicating that the magnetic structures were also well correlated on average, such that any static disorder of the magnetic structure must similarly affect all unit cells equally (i.e. local inhomogeneities are not prevalent).

By comparison, spin fluctuations can coexist with well correlated magnetic order \cite{1968Sakurai}, and have been shown to reduce moment magnitudes even at low temperature \cite{2005Janoschek}. The prevalence of spin fluctuations is emerging in the wider family of A-site columnar ordered quadruple perovskites that support competing exchange pathways. For example, in \ce{R2CuMnMn4O12} (R = Y or Dy), B site Mn$^{3+}$ moments appear to saturate at $\frac{3}{4}$ of their theoretical value, and only on cooling through a low temperature phase transition, proposed to originate in the softening of low energy magnons, do they recover their full moment \cite{2020Vibhakar}. 

\section{Conclusions}\label{SEC::conclusions}
All three magnetically dilute $x=1$, $x=2$ and $x=3$ \ce{Sm2MnMnMn_{4-x}Ti_xO_{12}} samples adopt long range magnetically ordered phases below 27 K, 62 K and 34 K, respectively. The Mn ions were empirically found to adopt a collinear ferrimagnetic structure upto a direction in the $ab$ plane, and we propose that the moment direction within the plane is likely determined by $f$-$d$ exchange interactions between the Mn and Sm sublattices. We show that the introduction of 50\% magnetic dilution onto the B-site sublattice gives rise to a ferrimagnetic compensation point and magnetization reversal. Percolation calculations demonstrated that long range magnetic order in the $x=3$ sample can only occur if it percolates via \emph{both} A-B and B-B exchange, hence demonstrating the importance of A-B exchange in the A-site columnar ordered quadruple perovskite manganites. Finally we showed that the unusual variation in transition temperature that occurred upon magnetic dilution was reflected in the reduction of ground state magnetic moments observed on all Mn sites and in all samples. Together, these results suggest the presence of spin fluctuations and or disorder leading to a departure from mean field physics. In future studies it would be interesting to perform inelastic neutron scattering experiments to  measure the Sm$^{3+}$ CEF energy levels and further refine our model of the magnetic anisotropy, as well as probe the presence of spin fluctuations at the lowest measured temperatures.

\section{Acknowledgements}
R. D. J. acknowledges financial support from the Royal Society. K. Y. and A. A. B. acknowledge JSPS KAKENHI Grant No. JP20H05276,
a research grant (40-37) from Nippon Sheet Glass
Foundation for Materials Science and Engineering, and
Innovative Science and Technology Initiative for Security
(Grant No. JPJ004596) from Acquisition, Technology, and
Logistics Agency (ATLA), Japan.

\bibliography{thesis}

%merlin.mbs apsrev4-1.bst 2010-07-25 4.21a (PWD, AO, DPC) hacked
%Control: key (0)
%Control: author (8) initials jnrlst
%Control: editor formatted (1) identically to author
%Control: production of article title (-1) disabled
%Control: page (0) single
%Control: year (1) truncated
%Control: production of eprint (0) enabled
\begin{thebibliography}{35}%
\makeatletter
\providecommand \@ifxundefined [1]{%
 \@ifx{#1\undefined}
}%
\providecommand \@ifnum [1]{%
 \ifnum #1\expandafter \@firstoftwo
 \else \expandafter \@secondoftwo
 \fi
}%
\providecommand \@ifx [1]{%
 \ifx #1\expandafter \@firstoftwo
 \else \expandafter \@secondoftwo
 \fi
}%
\providecommand \natexlab [1]{#1}%
\providecommand \enquote  [1]{``#1''}%
\providecommand \bibnamefont  [1]{#1}%
\providecommand \bibfnamefont [1]{#1}%
\providecommand \citenamefont [1]{#1}%
\providecommand \href@noop [0]{\@secondoftwo}%
\providecommand \href [0]{\begingroup \@sanitize@url \@href}%
\providecommand \@href[1]{\@@startlink{#1}\@@href}%
\providecommand \@@href[1]{\endgroup#1\@@endlink}%
\providecommand \@sanitize@url [0]{\catcode `\\12\catcode `\$12\catcode
  `\&12\catcode `\#12\catcode `\^12\catcode `\_12\catcode `\%12\relax}%
\providecommand \@@startlink[1]{}%
\providecommand \@@endlink[0]{}%
\providecommand \url  [0]{\begingroup\@sanitize@url \@url }%
\providecommand \@url [1]{\endgroup\@href {#1}{\urlprefix }}%
\providecommand \urlprefix  [0]{URL }%
\providecommand \Eprint [0]{\href }%
\providecommand \doibase [0]{http://dx.doi.org/}%
\providecommand \selectlanguage [0]{\@gobble}%
\providecommand \bibinfo  [0]{\@secondoftwo}%
\providecommand \bibfield  [0]{\@secondoftwo}%
\providecommand \translation [1]{[#1]}%
\providecommand \BibitemOpen [0]{}%
\providecommand \bibitemStop [0]{}%
\providecommand \bibitemNoStop [0]{.\EOS\space}%
\providecommand \EOS [0]{\spacefactor3000\relax}%
\providecommand \BibitemShut  [1]{\csname bibitem#1\endcsname}%
\let\auto@bib@innerbib\@empty
%</preamble>
\bibitem [{\citenamefont {Jin}\ \emph {et~al.}(1994)\citenamefont {Jin},
  \citenamefont {Tiefel}, \citenamefont {McCormack}, \citenamefont {Fastnacht},
  \citenamefont {Ramesh},\ and\ \citenamefont {Chen}}]{1994Jin}%
  \BibitemOpen
  \bibfield  {author} {\bibinfo {author} {\bibfnamefont {S.}~\bibnamefont
  {Jin}}, \bibinfo {author} {\bibfnamefont {T.~H.}\ \bibnamefont {Tiefel}},
  \bibinfo {author} {\bibfnamefont {M.}~\bibnamefont {McCormack}}, \bibinfo
  {author} {\bibfnamefont {R.}~\bibnamefont {Fastnacht}}, \bibinfo {author}
  {\bibfnamefont {R.}~\bibnamefont {Ramesh}}, \ and\ \bibinfo {author}
  {\bibfnamefont {L.}~\bibnamefont {Chen}},\ }\href@noop {} {\bibfield
  {journal} {\bibinfo  {journal} {Science}\ }\textbf {\bibinfo {volume}
  {264}},\ \bibinfo {pages} {413} (\bibinfo {year} {1994})}\BibitemShut
  {NoStop}%
\bibitem [{\citenamefont {Wollan}\ and\ \citenamefont
  {Koehler}(1955)}]{1955Wollan}%
  \BibitemOpen
  \bibfield  {author} {\bibinfo {author} {\bibfnamefont {E.~O.}\ \bibnamefont
  {Wollan}}\ and\ \bibinfo {author} {\bibfnamefont {W.~C.}\ \bibnamefont
  {Koehler}},\ }\href {\doibase 10.1103/PhysRev.100.545} {\bibfield  {journal}
  {\bibinfo  {journal} {Phys. Rev.}\ }\textbf {\bibinfo {volume} {100}},\
  \bibinfo {pages} {545} (\bibinfo {year} {1955})}\BibitemShut {NoStop}%
\bibitem [{\citenamefont {Zhou}\ and\ \citenamefont
  {Goodenough}(2006)}]{2006Zhou}%
  \BibitemOpen
  \bibfield  {author} {\bibinfo {author} {\bibfnamefont {J.-S.}\ \bibnamefont
  {Zhou}}\ and\ \bibinfo {author} {\bibfnamefont {J.~B.}\ \bibnamefont
  {Goodenough}},\ }\href {\doibase 10.1103/PhysRevLett.96.247202} {\bibfield
  {journal} {\bibinfo  {journal} {Phys. Rev. Lett.}\ }\textbf {\bibinfo
  {volume} {96}},\ \bibinfo {pages} {247202} (\bibinfo {year}
  {2006})}\BibitemShut {NoStop}%
\bibitem [{\citenamefont {Radaelli}\ \emph {et~al.}(1997)\citenamefont
  {Radaelli}, \citenamefont {Cox}, \citenamefont {Marezio},\ and\ \citenamefont
  {Cheong}}]{1997Radaelli}%
  \BibitemOpen
  \bibfield  {author} {\bibinfo {author} {\bibfnamefont {P.~G.}\ \bibnamefont
  {Radaelli}}, \bibinfo {author} {\bibfnamefont {D.~E.}\ \bibnamefont {Cox}},
  \bibinfo {author} {\bibfnamefont {M.}~\bibnamefont {Marezio}}, \ and\
  \bibinfo {author} {\bibfnamefont {S.-W.}\ \bibnamefont {Cheong}},\ }\href
  {\doibase 10.1103/PhysRevB.55.3015} {\bibfield  {journal} {\bibinfo
  {journal} {Phys. Rev. B}\ }\textbf {\bibinfo {volume} {55}},\ \bibinfo
  {pages} {3015} (\bibinfo {year} {1997})}\BibitemShut {NoStop}%
\bibitem [{\citenamefont {Goodenough}(1955)}]{1955Goodenough}%
  \BibitemOpen
  \bibfield  {author} {\bibinfo {author} {\bibfnamefont {J.~B.}\ \bibnamefont
  {Goodenough}},\ }\href {\doibase 10.1103/PhysRev.100.564} {\bibfield
  {journal} {\bibinfo  {journal} {Phys. Rev.}\ }\textbf {\bibinfo {volume}
  {100}},\ \bibinfo {pages} {564} (\bibinfo {year} {1955})}\BibitemShut
  {NoStop}%
\bibitem [{\citenamefont {Mu{\~n}oz}\ \emph {et~al.}(2001)\citenamefont
  {Mu{\~n}oz}, \citenamefont {Cas{\'a}is}, \citenamefont {Alonso},
  \citenamefont {Mart{\'\i}nez-Lope}, \citenamefont {Mart{\'\i}nez},\ and\
  \citenamefont {Fern{\'a}ndez-D{\'\i}az}}]{2001Munoz_orth}%
  \BibitemOpen
  \bibfield  {author} {\bibinfo {author} {\bibfnamefont {A.}~\bibnamefont
  {Mu{\~n}oz}}, \bibinfo {author} {\bibfnamefont {M.~T.}\ \bibnamefont
  {Cas{\'a}is}}, \bibinfo {author} {\bibfnamefont {J.~A.}\ \bibnamefont
  {Alonso}}, \bibinfo {author} {\bibfnamefont {M.~J.}\ \bibnamefont
  {Mart{\'\i}nez-Lope}}, \bibinfo {author} {\bibfnamefont {J.~L.}\ \bibnamefont
  {Mart{\'\i}nez}}, \ and\ \bibinfo {author} {\bibfnamefont {M.~T.}\
  \bibnamefont {Fern{\'a}ndez-D{\'\i}az}},\ }\href {\doibase 10.1021/ic0011009}
  {\bibfield  {journal} {\bibinfo  {journal} {Inorganic Chemistry}\ }\textbf
  {\bibinfo {volume} {40}},\ \bibinfo {pages} {1020} (\bibinfo {year}
  {2001})}\BibitemShut {NoStop}%
\bibitem [{\citenamefont {Glazer}(1972)}]{1972GlazerNotation}%
  \BibitemOpen
  \bibfield  {author} {\bibinfo {author} {\bibfnamefont {A.~M.}\ \bibnamefont
  {Glazer}},\ }\href {\doibase 10.1107/S0567740872007976} {\bibfield  {journal}
  {\bibinfo  {journal} {Acta Crystallographica Section B}\ }\textbf {\bibinfo
  {volume} {28}},\ \bibinfo {pages} {3384} (\bibinfo {year}
  {1972})}\BibitemShut {NoStop}%
\bibitem [{\citenamefont {Belik}\ \emph {et~al.}(2016)\citenamefont {Belik},
  \citenamefont {Glazkova}, \citenamefont {Katsuya}, \citenamefont {Tanaka},
  \citenamefont {Sobolev},\ and\ \citenamefont {Presniakov}}]{2016Belik}%
  \BibitemOpen
  \bibfield  {author} {\bibinfo {author} {\bibfnamefont {A.~A.}\ \bibnamefont
  {Belik}}, \bibinfo {author} {\bibfnamefont {Y.~S.}\ \bibnamefont {Glazkova}},
  \bibinfo {author} {\bibfnamefont {Y.}~\bibnamefont {Katsuya}}, \bibinfo
  {author} {\bibfnamefont {M.}~\bibnamefont {Tanaka}}, \bibinfo {author}
  {\bibfnamefont {A.~V.}\ \bibnamefont {Sobolev}}, \ and\ \bibinfo {author}
  {\bibfnamefont {I.~A.}\ \bibnamefont {Presniakov}},\ }\href@noop {}
  {\bibfield  {journal} {\bibinfo  {journal} {The Journal of Physical Chemistry
  C}\ }\textbf {\bibinfo {volume} {120}},\ \bibinfo {pages} {8278} (\bibinfo
  {year} {2016})}\BibitemShut {NoStop}%
\bibitem [{\citenamefont {Zhang}\ \emph {et~al.}(2018)\citenamefont {Zhang},
  \citenamefont {Terada}, \citenamefont {Johnson}, \citenamefont {Khalyavin},
  \citenamefont {Manuel}, \citenamefont {Katsuya}, \citenamefont {Tanaka},
  \citenamefont {Matsushita}, \citenamefont {Yamaura},\ and\ \citenamefont
  {Belik}}]{2018Zhang_RMn7O12}%
  \BibitemOpen
  \bibfield  {author} {\bibinfo {author} {\bibfnamefont {L.}~\bibnamefont
  {Zhang}}, \bibinfo {author} {\bibfnamefont {N.}~\bibnamefont {Terada}},
  \bibinfo {author} {\bibfnamefont {R.~D.}\ \bibnamefont {Johnson}}, \bibinfo
  {author} {\bibfnamefont {D.~D.}\ \bibnamefont {Khalyavin}}, \bibinfo {author}
  {\bibfnamefont {P.}~\bibnamefont {Manuel}}, \bibinfo {author} {\bibfnamefont
  {Y.}~\bibnamefont {Katsuya}}, \bibinfo {author} {\bibfnamefont
  {M.}~\bibnamefont {Tanaka}}, \bibinfo {author} {\bibfnamefont
  {Y.}~\bibnamefont {Matsushita}}, \bibinfo {author} {\bibfnamefont
  {K.}~\bibnamefont {Yamaura}}, \ and\ \bibinfo {author} {\bibfnamefont
  {A.~A.}\ \bibnamefont {Belik}},\ }\href {\doibase
  10.1021/acs.inorgchem.8b00479} {\bibfield  {journal} {\bibinfo  {journal}
  {Inorganic Chemistry}\ }\textbf {\bibinfo {volume} {57}},\ \bibinfo {pages}
  {5987} (\bibinfo {year} {2018})},\ \bibinfo {note} {pMID: 29722530},\ \Eprint
  {http://arxiv.org/abs/https://doi.org/10.1021/acs.inorgchem.8b00479}
  {https://doi.org/10.1021/acs.inorgchem.8b00479} \BibitemShut {NoStop}%
\bibitem [{\citenamefont {Johnson}\ \emph
  {et~al.}(2018{\natexlab{a}})\citenamefont {Johnson}, \citenamefont
  {Khalyavin}, \citenamefont {Manuel}, \citenamefont {Zhang}, \citenamefont
  {Yamaura},\ and\ \citenamefont {Belik}}]{2018Johnson_RMn7O12}%
  \BibitemOpen
  \bibfield  {author} {\bibinfo {author} {\bibfnamefont {R.~D.}\ \bibnamefont
  {Johnson}}, \bibinfo {author} {\bibfnamefont {D.~D.}\ \bibnamefont
  {Khalyavin}}, \bibinfo {author} {\bibfnamefont {P.}~\bibnamefont {Manuel}},
  \bibinfo {author} {\bibfnamefont {L.}~\bibnamefont {Zhang}}, \bibinfo
  {author} {\bibfnamefont {K.}~\bibnamefont {Yamaura}}, \ and\ \bibinfo
  {author} {\bibfnamefont {A.~A.}\ \bibnamefont {Belik}},\ }\href {\doibase
  10.1103/PhysRevB.98.104423} {\bibfield  {journal} {\bibinfo  {journal} {Phys.
  Rev. B}\ }\textbf {\bibinfo {volume} {98}},\ \bibinfo {pages} {104423}
  (\bibinfo {year} {2018}{\natexlab{a}})}\BibitemShut {NoStop}%
\bibitem [{\citenamefont {Johnson}\ \emph {et~al.}(2019)\citenamefont
  {Johnson}, \citenamefont {Khalyavin}, \citenamefont {Manuel}, \citenamefont
  {Katsuya}, \citenamefont {Tanaka}, \citenamefont {Matsushita}, \citenamefont
  {Zhang}, \citenamefont {Yamaura},\ and\ \citenamefont
  {Belik}}]{2019Johnson_YMn7O12}%
  \BibitemOpen
  \bibfield  {author} {\bibinfo {author} {\bibfnamefont {R.~D.}\ \bibnamefont
  {Johnson}}, \bibinfo {author} {\bibfnamefont {D.~D.}\ \bibnamefont
  {Khalyavin}}, \bibinfo {author} {\bibfnamefont {P.}~\bibnamefont {Manuel}},
  \bibinfo {author} {\bibfnamefont {Y.}~\bibnamefont {Katsuya}}, \bibinfo
  {author} {\bibfnamefont {M.}~\bibnamefont {Tanaka}}, \bibinfo {author}
  {\bibfnamefont {Y.}~\bibnamefont {Matsushita}}, \bibinfo {author}
  {\bibfnamefont {L.}~\bibnamefont {Zhang}}, \bibinfo {author} {\bibfnamefont
  {K.}~\bibnamefont {Yamaura}}, \ and\ \bibinfo {author} {\bibfnamefont
  {A.~A.}\ \bibnamefont {Belik}},\ }\href {\doibase 10.1103/PhysRevB.99.024107}
  {\bibfield  {journal} {\bibinfo  {journal} {Phys. Rev. B}\ }\textbf {\bibinfo
  {volume} {99}},\ \bibinfo {pages} {024107} (\bibinfo {year}
  {2019})}\BibitemShut {NoStop}%
\bibitem [{\citenamefont {Johnson}\ \emph {et~al.}(2016)\citenamefont
  {Johnson}, \citenamefont {Khalyavin}, \citenamefont {Manuel}, \citenamefont
  {Bombardi}, \citenamefont {Martin}, \citenamefont {Chapon},\ and\
  \citenamefont {Radaelli}}]{2016Johnson}%
  \BibitemOpen
  \bibfield  {author} {\bibinfo {author} {\bibfnamefont {R.~D.}\ \bibnamefont
  {Johnson}}, \bibinfo {author} {\bibfnamefont {D.~D.}\ \bibnamefont
  {Khalyavin}}, \bibinfo {author} {\bibfnamefont {P.}~\bibnamefont {Manuel}},
  \bibinfo {author} {\bibfnamefont {A.}~\bibnamefont {Bombardi}}, \bibinfo
  {author} {\bibfnamefont {C.}~\bibnamefont {Martin}}, \bibinfo {author}
  {\bibfnamefont {L.~C.}\ \bibnamefont {Chapon}}, \ and\ \bibinfo {author}
  {\bibfnamefont {P.~G.}\ \bibnamefont {Radaelli}},\ }\href {\doibase
  10.1103/PhysRevB.93.180403} {\bibfield  {journal} {\bibinfo  {journal} {Phys.
  Rev. B}\ }\textbf {\bibinfo {volume} {93}},\ \bibinfo {pages} {180403(R)}
  (\bibinfo {year} {2016})}\BibitemShut {NoStop}%
\bibitem [{\citenamefont {Johnson}\ \emph {et~al.}(2017)\citenamefont
  {Johnson}, \citenamefont {Khalyavin}, \citenamefont {Manuel}, \citenamefont
  {Radaelli}, \citenamefont {Glazkova}, \citenamefont {Terada},\ and\
  \citenamefont {Belik}}]{2017Johnson}%
  \BibitemOpen
  \bibfield  {author} {\bibinfo {author} {\bibfnamefont {R.~D.}\ \bibnamefont
  {Johnson}}, \bibinfo {author} {\bibfnamefont {D.~D.}\ \bibnamefont
  {Khalyavin}}, \bibinfo {author} {\bibfnamefont {P.}~\bibnamefont {Manuel}},
  \bibinfo {author} {\bibfnamefont {P.~G.}\ \bibnamefont {Radaelli}}, \bibinfo
  {author} {\bibfnamefont {I.~S.}\ \bibnamefont {Glazkova}}, \bibinfo {author}
  {\bibfnamefont {N.}~\bibnamefont {Terada}}, \ and\ \bibinfo {author}
  {\bibfnamefont {A.~A.}\ \bibnamefont {Belik}},\ }\href {\doibase
  10.1103/PhysRevB.96.054448} {\bibfield  {journal} {\bibinfo  {journal} {Phys.
  Rev. B}\ }\textbf {\bibinfo {volume} {96}},\ \bibinfo {pages} {054448}
  (\bibinfo {year} {2017})}\BibitemShut {NoStop}%
\bibitem [{\citenamefont {Johnson}\ \emph
  {et~al.}(2018{\natexlab{b}})\citenamefont {Johnson}, \citenamefont
  {Mezzadri}, \citenamefont {Manuel}, \citenamefont {Khalyavin}, \citenamefont
  {Gilioli},\ and\ \citenamefont {Radaelli}}]{2018Johnson_NaCaMn7O12}%
  \BibitemOpen
  \bibfield  {author} {\bibinfo {author} {\bibfnamefont {R.~D.}\ \bibnamefont
  {Johnson}}, \bibinfo {author} {\bibfnamefont {F.}~\bibnamefont {Mezzadri}},
  \bibinfo {author} {\bibfnamefont {P.}~\bibnamefont {Manuel}}, \bibinfo
  {author} {\bibfnamefont {D.~D.}\ \bibnamefont {Khalyavin}}, \bibinfo {author}
  {\bibfnamefont {E.}~\bibnamefont {Gilioli}}, \ and\ \bibinfo {author}
  {\bibfnamefont {P.~G.}\ \bibnamefont {Radaelli}},\ }\href {\doibase
  10.1103/PhysRevLett.120.257202} {\bibfield  {journal} {\bibinfo  {journal}
  {Phys. Rev. Lett.}\ }\textbf {\bibinfo {volume} {120}},\ \bibinfo {pages}
  {257202} (\bibinfo {year} {2018}{\natexlab{b}})}\BibitemShut {NoStop}%
\bibitem [{\citenamefont {Zhang}\ \emph {et~al.}(2017)\citenamefont {Zhang},
  \citenamefont {Matsushita}, \citenamefont {Yamaura},\ and\ \citenamefont
  {Belik}}]{2017Zhang}%
  \BibitemOpen
  \bibfield  {author} {\bibinfo {author} {\bibfnamefont {L.}~\bibnamefont
  {Zhang}}, \bibinfo {author} {\bibfnamefont {Y.}~\bibnamefont {Matsushita}},
  \bibinfo {author} {\bibfnamefont {K.}~\bibnamefont {Yamaura}}, \ and\
  \bibinfo {author} {\bibfnamefont {A.~A.}\ \bibnamefont {Belik}},\ }\href
  {\doibase 10.1021/acs.inorgchem.7b00347} {\bibfield  {journal} {\bibinfo
  {journal} {Inorganic Chemistry}\ }\textbf {\bibinfo {volume} {56}},\ \bibinfo
  {pages} {5210} (\bibinfo {year} {2017})},\ \bibinfo {note} {pMID: 28425715},\
  \Eprint {http://arxiv.org/abs/https://doi.org/10.1021/acs.inorgchem.7b00347}
  {https://doi.org/10.1021/acs.inorgchem.7b00347} \BibitemShut {NoStop}%
\bibitem [{\citenamefont {Liu}\ \emph {et~al.}(2020)\citenamefont {Liu},
  \citenamefont {Scatena}, \citenamefont {Khalyavin}, \citenamefont {Johnson},
  \citenamefont {Inaguma}, \citenamefont {Tanaka}, \citenamefont {Matsushita},
  \citenamefont {Yamaura},\ and\ \citenamefont {Belik}}]{2020Liu}%
  \BibitemOpen
  \bibfield  {author} {\bibinfo {author} {\bibfnamefont {R.}~\bibnamefont
  {Liu}}, \bibinfo {author} {\bibfnamefont {R.}~\bibnamefont {Scatena}},
  \bibinfo {author} {\bibfnamefont {D.~D.}\ \bibnamefont {Khalyavin}}, \bibinfo
  {author} {\bibfnamefont {R.~D.}\ \bibnamefont {Johnson}}, \bibinfo {author}
  {\bibfnamefont {Y.}~\bibnamefont {Inaguma}}, \bibinfo {author} {\bibfnamefont
  {M.}~\bibnamefont {Tanaka}}, \bibinfo {author} {\bibfnamefont
  {Y.}~\bibnamefont {Matsushita}}, \bibinfo {author} {\bibfnamefont
  {K.}~\bibnamefont {Yamaura}}, \ and\ \bibinfo {author} {\bibfnamefont
  {A.~A.}\ \bibnamefont {Belik}},\ }\href {\doibase
  10.1021/acs.inorgchem.0c00938} {\bibfield  {journal} {\bibinfo  {journal}
  {Inorganic Chemistry}\ }\textbf {\bibinfo {volume} {59}},\ \bibinfo {pages}
  {9065} (\bibinfo {year} {2020})}\BibitemShut {NoStop}%
\bibitem [{\citenamefont {Vibhakar}\ \emph {et~al.}(2019)\citenamefont
  {Vibhakar}, \citenamefont {Khalyavin}, \citenamefont {Manuel}, \citenamefont
  {Zhang}, \citenamefont {Yamaura}, \citenamefont {Radaelli}, \citenamefont
  {Belik},\ and\ \citenamefont {Johnson}}]{2019Vibhakar}%
  \BibitemOpen
  \bibfield  {author} {\bibinfo {author} {\bibfnamefont {A.~M.}\ \bibnamefont
  {Vibhakar}}, \bibinfo {author} {\bibfnamefont {D.~D.}\ \bibnamefont
  {Khalyavin}}, \bibinfo {author} {\bibfnamefont {P.}~\bibnamefont {Manuel}},
  \bibinfo {author} {\bibfnamefont {L.}~\bibnamefont {Zhang}}, \bibinfo
  {author} {\bibfnamefont {K.}~\bibnamefont {Yamaura}}, \bibinfo {author}
  {\bibfnamefont {P.~G.}\ \bibnamefont {Radaelli}}, \bibinfo {author}
  {\bibfnamefont {A.~A.}\ \bibnamefont {Belik}}, \ and\ \bibinfo {author}
  {\bibfnamefont {R.~D.}\ \bibnamefont {Johnson}},\ }\href {\doibase
  10.1103/PhysRevB.99.104424} {\bibfield  {journal} {\bibinfo  {journal} {Phys.
  Rev. B}\ }\textbf {\bibinfo {volume} {99}},\ \bibinfo {pages} {104424}
  (\bibinfo {year} {2019})}\BibitemShut {NoStop}%
\bibitem [{\citenamefont {Vibhakar}\ \emph {et~al.}(2020)\citenamefont
  {Vibhakar}, \citenamefont {Khalyavin}, \citenamefont {Manuel}, \citenamefont
  {Liu}, \citenamefont {Belik},\ and\ \citenamefont {Johnson}}]{2020Vibhakar}%
  \BibitemOpen
  \bibfield  {author} {\bibinfo {author} {\bibfnamefont {A.~M.}\ \bibnamefont
  {Vibhakar}}, \bibinfo {author} {\bibfnamefont {D.~D.}\ \bibnamefont
  {Khalyavin}}, \bibinfo {author} {\bibfnamefont {P.}~\bibnamefont {Manuel}},
  \bibinfo {author} {\bibfnamefont {J.}~\bibnamefont {Liu}}, \bibinfo {author}
  {\bibfnamefont {A.~A.}\ \bibnamefont {Belik}}, \ and\ \bibinfo {author}
  {\bibfnamefont {R.~D.}\ \bibnamefont {Johnson}},\ }\href {\doibase
  10.1103/PhysRevLett.124.127201} {\bibfield  {journal} {\bibinfo  {journal}
  {Phys. Rev. Lett.}\ }\textbf {\bibinfo {volume} {124}},\ \bibinfo {pages}
  {127201} (\bibinfo {year} {2020})}\BibitemShut {NoStop}%
\bibitem [{\citenamefont {Belik}\ \emph {et~al.}(2019)\citenamefont {Belik},
  \citenamefont {Zhang}, \citenamefont {Liu}, \citenamefont {Khalyavin},
  \citenamefont {Katsuya}, \citenamefont {Tanaka},\ and\ \citenamefont
  {Yamaura}}]{2019Belik}%
  \BibitemOpen
  \bibfield  {author} {\bibinfo {author} {\bibfnamefont {A.~A.}\ \bibnamefont
  {Belik}}, \bibinfo {author} {\bibfnamefont {L.}~\bibnamefont {Zhang}},
  \bibinfo {author} {\bibfnamefont {R.}~\bibnamefont {Liu}}, \bibinfo {author}
  {\bibfnamefont {D.~D.}\ \bibnamefont {Khalyavin}}, \bibinfo {author}
  {\bibfnamefont {Y.}~\bibnamefont {Katsuya}}, \bibinfo {author} {\bibfnamefont
  {M.}~\bibnamefont {Tanaka}}, \ and\ \bibinfo {author} {\bibfnamefont
  {K.}~\bibnamefont {Yamaura}},\ }\href {\doibase
  10.1021/acs.inorgchem.9b00049} {\bibfield  {journal} {\bibinfo  {journal}
  {Inorganic Chemistry}\ }\textbf {\bibinfo {volume} {58}},\ \bibinfo {pages}
  {3492} (\bibinfo {year} {2019})},\ \Eprint
  {http://arxiv.org/abs/https://doi.org/10.1021/acs.inorgchem.9b00049}
  {https://doi.org/10.1021/acs.inorgchem.9b00049} \BibitemShut {NoStop}%
\bibitem [{\citenamefont {Elliott}\ and\ \citenamefont
  {Heap}(1962)}]{1962Elliott}%
  \BibitemOpen
  \bibfield  {author} {\bibinfo {author} {\bibfnamefont {R.~J.}\ \bibnamefont
  {Elliott}}\ and\ \bibinfo {author} {\bibfnamefont {B.}~\bibnamefont {Heap}},\
  }\href@noop {} {\bibfield  {journal} {\bibinfo  {journal} {Proceedings of the
  Royal Society of London. Series A. Mathematical and Physical Sciences}\
  }\textbf {\bibinfo {volume} {265}},\ \bibinfo {pages} {264} (\bibinfo {year}
  {1962})}\BibitemShut {NoStop}%
\bibitem [{\citenamefont {Chapon}\ \emph {et~al.}(2011)\citenamefont {Chapon},
  \citenamefont {Manuel}, \citenamefont {Radaelli}, \citenamefont {Benson},
  \citenamefont {Perrott}, \citenamefont {Ansell}, \citenamefont {Rhodes},
  \citenamefont {Raspino}, \citenamefont {Duxbury}, \citenamefont {Spill},\
  and\ \citenamefont {Norris}}]{2011ChaponWISH}%
  \BibitemOpen
  \bibfield  {author} {\bibinfo {author} {\bibfnamefont {L.~C.}\ \bibnamefont
  {Chapon}}, \bibinfo {author} {\bibfnamefont {P.}~\bibnamefont {Manuel}},
  \bibinfo {author} {\bibfnamefont {P.~G.}\ \bibnamefont {Radaelli}}, \bibinfo
  {author} {\bibfnamefont {C.}~\bibnamefont {Benson}}, \bibinfo {author}
  {\bibfnamefont {L.}~\bibnamefont {Perrott}}, \bibinfo {author} {\bibfnamefont
  {S.}~\bibnamefont {Ansell}}, \bibinfo {author} {\bibfnamefont {N.~J.}\
  \bibnamefont {Rhodes}}, \bibinfo {author} {\bibfnamefont {D.}~\bibnamefont
  {Raspino}}, \bibinfo {author} {\bibfnamefont {D.}~\bibnamefont {Duxbury}},
  \bibinfo {author} {\bibfnamefont {E.}~\bibnamefont {Spill}}, \ and\ \bibinfo
  {author} {\bibfnamefont {J.}~\bibnamefont {Norris}},\ }\href {\doibase
  10.1080/10448632.2011.569650} {\bibfield  {journal} {\bibinfo  {journal}
  {Neutron News}\ }\textbf {\bibinfo {volume} {22}},\ \bibinfo {pages} {22}
  (\bibinfo {year} {2011})},\ \Eprint
  {http://arxiv.org/abs/https://doi.org/10.1080/10448632.2011.569650}
  {https://doi.org/10.1080/10448632.2011.569650} \BibitemShut {NoStop}%
\bibitem [{\citenamefont {Campbell}\ \emph {et~al.}(2006)\citenamefont
  {Campbell}, \citenamefont {Stokes}, \citenamefont {Tanner},\ and\
  \citenamefont {Hatch}}]{2006Isodistort}%
  \BibitemOpen
  \bibfield  {author} {\bibinfo {author} {\bibfnamefont {B.~J.}\ \bibnamefont
  {Campbell}}, \bibinfo {author} {\bibfnamefont {H.~T.}\ \bibnamefont
  {Stokes}}, \bibinfo {author} {\bibfnamefont {D.~E.}\ \bibnamefont {Tanner}},
  \ and\ \bibinfo {author} {\bibfnamefont {D.~M.}\ \bibnamefont {Hatch}},\
  }\href {\doibase 10.1107/S0021889806014075} {\bibfield  {journal} {\bibinfo
  {journal} {Journal of Applied Crystallography}\ }\textbf {\bibinfo {volume}
  {39}},\ \bibinfo {pages} {607} (\bibinfo {year} {2006})}\BibitemShut
  {NoStop}%
\bibitem [{\citenamefont {Rodr{\'\i}guez-Carvajal}(1993)}]{1993Rodriguez}%
  \BibitemOpen
  \bibfield  {author} {\bibinfo {author} {\bibfnamefont {J.}~\bibnamefont
  {Rodr{\'\i}guez-Carvajal}},\ }\href {\doibase
  https://doi.org/10.1016/0921-4526(93)90108-I} {\bibfield  {journal} {\bibinfo
   {journal} {Physica B: Condensed Matter}\ }\textbf {\bibinfo {volume}
  {192}},\ \bibinfo {pages} {55 } (\bibinfo {year} {1993})}\BibitemShut
  {NoStop}%
\bibitem [{\citenamefont {Whangbo}\ \emph {et~al.}(2015)\citenamefont
  {Whangbo}, \citenamefont {Gordon}, \citenamefont {Xiang}, \citenamefont
  {Koo},\ and\ \citenamefont {Lee}}]{2015Whangbo}%
  \BibitemOpen
  \bibfield  {author} {\bibinfo {author} {\bibfnamefont {M.-H.}\ \bibnamefont
  {Whangbo}}, \bibinfo {author} {\bibfnamefont {E.~E.}\ \bibnamefont {Gordon}},
  \bibinfo {author} {\bibfnamefont {H.}~\bibnamefont {Xiang}}, \bibinfo
  {author} {\bibfnamefont {H.-J.}\ \bibnamefont {Koo}}, \ and\ \bibinfo
  {author} {\bibfnamefont {C.}~\bibnamefont {Lee}},\ }\href {\doibase
  10.1021/acs.accounts.5b00408} {\bibfield  {journal} {\bibinfo  {journal}
  {Accounts of Chemical Research}\ }\textbf {\bibinfo {volume} {48}},\ \bibinfo
  {pages} {3080} (\bibinfo {year} {2015})},\ \bibinfo {note} {pMID: 26616364},\
  \Eprint {http://arxiv.org/abs/https://doi.org/10.1021/acs.accounts.5b00408}
  {https://doi.org/10.1021/acs.accounts.5b00408} \BibitemShut {NoStop}%
\bibitem [{\citenamefont {White}(1969)}]{1969White}%
  \BibitemOpen
  \bibfield  {author} {\bibinfo {author} {\bibfnamefont {R.~L.}\ \bibnamefont
  {White}},\ }\href {\doibase 10.1063/1.1657530} {\bibfield  {journal}
  {\bibinfo  {journal} {Journal of Applied Physics}\ }\textbf {\bibinfo
  {volume} {40}},\ \bibinfo {pages} {1061} (\bibinfo {year} {1969})},\ \Eprint
  {http://arxiv.org/abs/https://doi.org/10.1063/1.1657530}
  {https://doi.org/10.1063/1.1657530} \BibitemShut {NoStop}%
\bibitem [{\citenamefont {Hutchings}(1964)}]{1964Hutchings}%
  \BibitemOpen
  \bibfield  {author} {\bibinfo {author} {\bibfnamefont {M.}~\bibnamefont
  {Hutchings}}\ }(\bibinfo  {publisher} {Academic Press},\ \bibinfo {year}
  {1964})\ pp.\ \bibinfo {pages} {227 -- 273}\BibitemShut {NoStop}%
\bibitem [{\citenamefont {Stauffer}\ and\ \citenamefont
  {Aharony}(1985)}]{1985Stauffer}%
  \BibitemOpen
  \bibfield  {author} {\bibinfo {author} {\bibfnamefont {D.}~\bibnamefont
  {Stauffer}}\ and\ \bibinfo {author} {\bibfnamefont {A.}~\bibnamefont
  {Aharony}},\ }\href@noop {} {\emph {\bibinfo {title} {Introduction to
  Percolation Theory}}},\ \bibinfo {edition} {2nd}\ ed.\ (\bibinfo  {publisher}
  {Taylor \& Francis Ltd},\ \bibinfo {year} {1985})\BibitemShut {NoStop}%
\bibitem [{\citenamefont {Landau}(1980)}]{1980Landau}%
  \BibitemOpen
  \bibfield  {author} {\bibinfo {author} {\bibfnamefont {D.~P.}\ \bibnamefont
  {Landau}},\ }\href {\doibase 10.1103/PhysRevB.22.2450} {\bibfield  {journal}
  {\bibinfo  {journal} {Phys. Rev. B}\ }\textbf {\bibinfo {volume} {22}},\
  \bibinfo {pages} {2450} (\bibinfo {year} {1980})}\BibitemShut {NoStop}%
\bibitem [{\citenamefont {Zvyagin}\ and\ \citenamefont
  {Anders}(1975)}]{1975Zvyagin}%
  \BibitemOpen
  \bibfield  {author} {\bibinfo {author} {\bibfnamefont {A.}~\bibnamefont
  {Zvyagin}}\ and\ \bibinfo {author} {\bibfnamefont {A.}~\bibnamefont
  {Anders}},\ }\href@noop {} {\bibfield  {journal} {\bibinfo  {journal} {JETP}\
  }\textbf {\bibinfo {volume} {40}},\ \bibinfo {pages} {154} (\bibinfo {year}
  {1975})}\BibitemShut {NoStop}%
\bibitem [{\citenamefont {Cheong}\ \emph {et~al.}(1991)\citenamefont {Cheong},
  \citenamefont {Cooper}, \citenamefont {Rupp}, \citenamefont {Batlogg},
  \citenamefont {Thompson},\ and\ \citenamefont {Fisk}}]{1991Cheong}%
  \BibitemOpen
  \bibfield  {author} {\bibinfo {author} {\bibfnamefont {S.-W.}\ \bibnamefont
  {Cheong}}, \bibinfo {author} {\bibfnamefont {A.~S.}\ \bibnamefont {Cooper}},
  \bibinfo {author} {\bibfnamefont {L.~W.}\ \bibnamefont {Rupp}}, \bibinfo
  {author} {\bibfnamefont {B.}~\bibnamefont {Batlogg}}, \bibinfo {author}
  {\bibfnamefont {J.~D.}\ \bibnamefont {Thompson}}, \ and\ \bibinfo {author}
  {\bibfnamefont {Z.}~\bibnamefont {Fisk}},\ }\href {\doibase
  10.1103/PhysRevB.44.9739} {\bibfield  {journal} {\bibinfo  {journal} {Phys.
  Rev. B}\ }\textbf {\bibinfo {volume} {44}},\ \bibinfo {pages} {9739}
  (\bibinfo {year} {1991})}\BibitemShut {NoStop}%
\bibitem [{\citenamefont {Niven}\ \emph {et~al.}(2014)\citenamefont {Niven},
  \citenamefont {Johnson}, \citenamefont {Bourque}, \citenamefont {Murray},
  \citenamefont {James}, \citenamefont {Dabkowska}, \citenamefont {Gaulin},\
  and\ \citenamefont {White}}]{2014Niven}%
  \BibitemOpen
  \bibfield  {author} {\bibinfo {author} {\bibfnamefont {J.~F.}\ \bibnamefont
  {Niven}}, \bibinfo {author} {\bibfnamefont {M.~B.}\ \bibnamefont {Johnson}},
  \bibinfo {author} {\bibfnamefont {A.}~\bibnamefont {Bourque}}, \bibinfo
  {author} {\bibfnamefont {P.~J.}\ \bibnamefont {Murray}}, \bibinfo {author}
  {\bibfnamefont {D.~D.}\ \bibnamefont {James}}, \bibinfo {author}
  {\bibfnamefont {H.~A.}\ \bibnamefont {Dabkowska}}, \bibinfo {author}
  {\bibfnamefont {B.~D.}\ \bibnamefont {Gaulin}}, \ and\ \bibinfo {author}
  {\bibfnamefont {M.~A.}\ \bibnamefont {White}},\ }\href@noop {} {\bibfield
  {journal} {\bibinfo  {journal} {Proceedings of the Royal Society A:
  Mathematical, Physical and Engineering Sciences}\ }\textbf {\bibinfo {volume}
  {470}},\ \bibinfo {pages} {20140387} (\bibinfo {year} {2014})}\BibitemShut
  {NoStop}%
\bibitem [{\citenamefont {Van~Vleck}(1945)}]{1945VanVleck}%
  \BibitemOpen
  \bibfield  {author} {\bibinfo {author} {\bibfnamefont {J.~H.}\ \bibnamefont
  {Van~Vleck}},\ }\href {\doibase 10.1103/RevModPhys.17.27} {\bibfield
  {journal} {\bibinfo  {journal} {Rev. Mod. Phys.}\ }\textbf {\bibinfo {volume}
  {17}},\ \bibinfo {pages} {27} (\bibinfo {year} {1945})}\BibitemShut {NoStop}%
\bibitem [{\citenamefont {Sakurai}\ \emph {et~al.}(1968)\citenamefont
  {Sakurai}, \citenamefont {Buyers}, \citenamefont {Cowley},\ and\
  \citenamefont {Dolling}}]{1968Sakurai}%
  \BibitemOpen
  \bibfield  {author} {\bibinfo {author} {\bibfnamefont {J.}~\bibnamefont
  {Sakurai}}, \bibinfo {author} {\bibfnamefont {W.~J.~L.}\ \bibnamefont
  {Buyers}}, \bibinfo {author} {\bibfnamefont {R.~A.}\ \bibnamefont {Cowley}},
  \ and\ \bibinfo {author} {\bibfnamefont {G.}~\bibnamefont {Dolling}},\ }\href
  {\doibase 10.1103/PhysRev.167.510} {\bibfield  {journal} {\bibinfo  {journal}
  {Phys. Rev.}\ }\textbf {\bibinfo {volume} {167}},\ \bibinfo {pages} {510}
  (\bibinfo {year} {1968})}\BibitemShut {NoStop}%
\bibitem [{\citenamefont {Janoschek}\ \emph {et~al.}(2005)\citenamefont
  {Janoschek}, \citenamefont {Roessli}, \citenamefont {Keller}, \citenamefont
  {Gvasaliya}, \citenamefont {Conder},\ and\ \citenamefont
  {Pomjakushina}}]{2005Janoschek}%
  \BibitemOpen
  \bibfield  {author} {\bibinfo {author} {\bibfnamefont {M.}~\bibnamefont
  {Janoschek}}, \bibinfo {author} {\bibfnamefont {B.}~\bibnamefont {Roessli}},
  \bibinfo {author} {\bibfnamefont {L.}~\bibnamefont {Keller}}, \bibinfo
  {author} {\bibfnamefont {S.~N.}\ \bibnamefont {Gvasaliya}}, \bibinfo {author}
  {\bibfnamefont {K.}~\bibnamefont {Conder}}, \ and\ \bibinfo {author}
  {\bibfnamefont {E.}~\bibnamefont {Pomjakushina}},\ }\href {\doibase
  10.1088/0953-8984/17/42/l01} {\bibfield  {journal} {\bibinfo  {journal}
  {Journal of Physics: Condensed Matter}\ }\textbf {\bibinfo {volume} {17}},\
  \bibinfo {pages} {L425} (\bibinfo {year} {2005})}\BibitemShut {NoStop}%
\bibitem [{\citenamefont {Brese}\ and\ \citenamefont
  {O'Keeffe}(1991)}]{1991Brese}%
  \BibitemOpen
  \bibfield  {author} {\bibinfo {author} {\bibfnamefont {N.~E.}\ \bibnamefont
  {Brese}}\ and\ \bibinfo {author} {\bibfnamefont {M.}~\bibnamefont
  {O'Keeffe}},\ }\href {\doibase 10.1107/S0108768190011041} {\bibfield
  {journal} {\bibinfo  {journal} {Acta Crystallographica Section B}\ }\textbf
  {\bibinfo {volume} {47}},\ \bibinfo {pages} {192} (\bibinfo {year} {1991})},\
  \Eprint
  {http://arxiv.org/abs/https://onlinelibrary.wiley.com/doi/pdf/10.1107/S0108768190011041}
  {https://onlinelibrary.wiley.com/doi/pdf/10.1107/S0108768190011041}
  \BibitemShut {NoStop}%
\end{thebibliography}%

\appendix

\section{Irreducible representations and their symmetry adapted basis functions \label{SEC::irreps}}

We give the basis functions of the $\Gamma_5^+$ irreducible representation used to describe the magnetic structures of \ce{Sm2MnMnMn_{4-x}Ti_xO_{12}} for the $A$, $A'$ and $A''$, and B sites in Tables \ref{TAB::Asites}, \ref{TAB::A'A''sites} and \ref{TAB::Bsites} respectively. 

\begin{table*}[!h]
\caption{\label{TAB::Asites} The $\Gamma_5^+$ symmetry adapted basis functions given for the $A$ site Wyckoff position in \ce{Sm2MnMnMn_{4-x}Ti_xO_{12}}, listed separately for the (a,0), (a,a) and (a,b) order parameter directions. Moment components of symmetry-equivalent ions have the same magnitude. Note that the modes and irreps are listed according to the labelling scheme adopted in the  \textsc{isotropy} software suite \cite{2006Isodistort}.}
\begin{ruledtabular}
\begin{tabular}{c c | c c c c}
O. P. & Mode & 0.25, 0.25, $ z$ & 0.75, 0.75, $-z$ & 0.25, 0.25, $z + \frac{1}{2}$ & 0.75, 0.75, $ -z  + \frac{1}{2}$ \\
\hline
(a,0)&B2(a)  & [1, 0, 0]        & [1, 0, 0]        & [0, 0, 0]        & [0, 0, 0]          \\
     &B1(a)  & [0, 0, 0]        & [0, 0, 0]        & [1, 0, 0]        & [1, 0, 0]          \\
\hline
(a,a)&B2(a)   & [1, 0, 0]        & [1, 0, 0]        & [0, 1, 0]        & [0, 1, 0]       \\
     &B1(a)   &    [0, 1, 0]        & [0, 1, 0]        & [1, 0, 0]        & [1, 0, 0]       \\
\hline
(a,b)&B2(a)    & [1, 0, 0]        & [1, 0, 0]        & [0, 0, 0]        & [0, 0, 0]          \\
     &B2(b)    & [0, 0, 0]        & [0, 0, 0]        & [0, 1, 0]        & [0, 1, 0]           \\ 
     &B1(a)    & [0, 0, 0]        & [0, 0, 0]        & [1, 0, 0]        & [1, 0, 0]   \\ 
     &B1(b)    & [0, 1, 0]        & [0, 1, 0]        & [0, 0, 0]        & [0, 0, 0]                \\ 
\end{tabular}
\end{ruledtabular}
\end{table*}

\begin{table*}[!h]
\caption{\label{TAB::A'A''sites}The $\Gamma_5^+$ symmetry adapted basis functions given for the $A'$ and $A''$  site Wyckoff positions in \ce{Sm2MnMnMn_{4-x}Ti_xO_{12}}, listed separately for the (a,0), (a,a) and (a,b) order parameter directions. Here we use the Wyckoff position 2a of the $A'$ site, such that it is constrained to the centre of the square planar coordination. Moment components of symmetry-equivalent ions have the same magnitude. Note that the modes and irreps are listed according to the labelling scheme adopted in the \textsc{isotropy} software suite \cite{2006Isodistort}.}
\begin{ruledtabular}
\begin{tabular}{c c | c c}
O. P. & Mode       & 0.75, 0.25, 0.75 & 0.25, 0.75, 0.25\\
\hline
(a,0) & E(a)   & [1,0,0]               & [1,0,0]           \\
\hline
(a,a) & E(a)   & [1,1,0]               & [1,1,0]           \\
\hline
(a,b) & E(a)   & [1,0,0]               & [1,0,0]           \\
      & E(b)   & [0,1,0]               & [0,1,0]           \\
\end{tabular}
\end{ruledtabular}
\end{table*}

\begin{table*}[!h]
\caption{\label{TAB::Bsites}The $\Gamma_5^+$ symmetry adapted basis functions given for the B site Wyckoff position in \ce{Sm2MnMnMn_{4-x}Ti_xO_{12}}, listed separately for the (a,0), (a,a) and (a,b) order parameter directions. Moment components of symmetry-equivalent ions have the same magnitude. Note that the modes and irreps are listed according to the labelling scheme adopted in the \textsc{isotropy} software suite \cite{2006Isodistort}.}
\begin{ruledtabular}
\begin{tabular}{c c | c c c c c c c c}
O. P. & Mode       & 0.0, 0.0, 0.0 & 0.5, 0.0, 0.0 & 0.0, 0.5, 0.0 & 0.5, 0.5, 0.0 & 0.0, 0.0, 0.5 & 0.5, 0.0, 0.5 & 0.0, 0.5, 0.5 &  0.5, 0.5, 0.5\\
\hline 
(a,0)     & $\mathrm{Ag_1(a)}$ & [1, 0, 0] & [1, 0,  0] & [1, 0,  0] & [1, 0, 0]  & [0, 0, 0] & [0, 0, 0]  & [0, 0, 0]  & [0, 0, 0] \\
          & $\mathrm{Ag_2(a)}$ & [0, 0, 0] & [0, 0, 0]  & [0, 0, 0]  & [0, 0, 0]  & [0, 1, 0] & [0, -1, 0] & [0, -1, 0] & [0, 1, 0] \\
          & $\mathrm{Ag_3(a)}$ & [0, 1, 0] & [0, -1, 0] & [0, -1, 0] & [0, 1, 0]  & [0, 0, 0] & [0, 0, 0]  & [0, 0, 0]  & [0, 0, 0] \\
          & $\mathrm{Ag_4(a)}$ & [0, 0, 0] & [0, 0, 0]  & [0, 0, 0]  & [0, 0, 0]  & [1, 0, 0] & [1, 0, 0]  & [1, 0, 0]  & [1, 0, 0] \\
          & $\mathrm{Ag_5(a)}$ & [0, 0, 1] & [0, 0, -1] & [0, 0, 1]  & [0, 0, -1] & [0, 0, 0] & [0, 0, 0]  & [0, 0, 0]  & [0, 0, 0]\\
          & $\mathrm{Ag_6(a)}$ & [0, 0, 0] & [0, 0, 0]  & [0, 0, 0]  & [0, 0, 0]  & [0, 0, 1] & [0, 0, -1] & [0, 0, 1] & [0, 0, -1] \\
\hline
(a,a)     & $\mathrm{Ag_1(a)}$  & [1, 0, 0] & [1, 0, 0]  & [1, 0, 0]  & [1, 0, 0] & [0, 1, 0] & [0, 1, 0]  & [0, 1, 0]  & [0, 1, 0] \\
          & $\mathrm{Ag_2(a)}$  & [1, 0, 0] & [-1, 0, 0] & [-1, 0, 0] & [1, 0, 0] & [0, 1, 0] & [0, -1, 0] & [0, -1, 0] & [0, 1, 0] \\
          & $\mathrm{Ag_3(a)}$  & [0, 1, 0] & [0, -1, 0] & [0, -1, 0] & [0, 1, 0] & [1, 0, 0] & [-1, 0, 0] & [-1, 0, 0] & [1, 0, 0] \\
          & $\mathrm{Ag_4(a)}$  & [0, 1, 0] & [0, 1, 0]  & [0, 1, 0]  & [0, 1, 0] & [1, 0, 0] & [1, 0, 0]  & [1, 0, 0]  & [1, 0, 0]\\
          & $\mathrm{Ag_5(a)}$  & [0, 0, 1] & [0, 0, -1] & [0, 0, 1]  & [0, 0, -1] & [0, 0, 1] & [0, 0, 1]  & [0, 0, -1] & [0, 0, -1] \\
          & $\mathrm{Ag_6(a)}$  & [0, 0, 1] & [0, 0, 1]  & [0, 0, -1] & [0, 0, -1] & [0, 0, 1] & [0, 0, -1] & [0, 0, 1]  & [0, 0, -1] \\
\hline
(a,b)     & $\mathrm{Ag_1(a)}$  & [1, 0, 0] & [1, 0, 0]  & [1, 0, 0]  & [1, 0, 0] &  [0, 0, 0] &  [0, 0, 0] &  [0, 0, 0] &  [0, 0, 0] \\
          & $\mathrm{Ag_1(b)}$  & [0, 0, 0] & [0, 0, 0]  & [0, 0, 0]  & [0, 0, 0] &  [0, 1, 0] &  [0, 1, 0] &  [0, 1, 0] &  [0, 1, 0] \\
          & $\mathrm{Ag_2(a)}$  & [0, 0, 0] & [0, 0, 0]  & [0, 0, 0]  & [0, 0, 0] &  [0, 1, 0] &  [0, -1, 0] & [0, -1, 0] & [0, 1, 0] \\
          & $\mathrm{Ag_2(b)}$  & [1, 0, 0] & [-1, 0, 0] & [-1, 0, 0] & [1, 0, 0] &  [0, 0, 0] &  [0, 0, 0] &  [0, 0, 0] &  [0, 0, 0] \\
          & $\mathrm{Ag_3(a)}$  & [0, 1, 0] & [0, -1, 0] & [0, -1, 0] & [0, 1, 0] &  [0, 0, 0] &  [0, 0, 0] &  [0, 0, 0] &  [0, 0, 0] \\
          & $\mathrm{Ag_3(b)}$  & [0, 0, 0] & [0, 0, 0]  & [0, 0, 0]  & [0, 0, 0] & [1, 0, 0] & [-1, 0, 0] & [-1, 0, 0] & [1, 0, 0]  \\
          & $\mathrm{Ag_4(a)}$  & [0, 0, 0] & [0, 0, 0]  & [0, 0, 0]  & [0, 0, 0] & [1, 0, 0] & [1, 0, 0]  & [1, 0, 0]  & [1, 0, 0] \\
          & $\mathrm{Ag_4(b)}$  & [0, 1, 0] & [0, 1, 0]  & [0, 1, 0]  & [0, 1, 0] &  [0, 0, 0] &  [0, 0, 0] &  [0, 0, 0] &  [0, 0, 0]\\
          & $\mathrm{Ag_5(a)}$  & [0, 0, 1] & [0, 0, -1] & [0, 0, 1]  & [0, 0, -1]&  [0, 0, 0] &  [0, 0, 0] &  [0, 0, 0] &  [0, 0, 0]  \\
          & $\mathrm{Ag_5(b)}$  & [0, 0, 0] & [0, 0, 0]  & [0, 0, 0]  & [0, 0, 0] & [0, 0, 1] & [0, 0, 1] & [0, 0, -1] & [0, 0, -1] \\
          & $\mathrm{Ag_6(a)}$  & [0, 0, 0] & [0, 0, 0]  & [0, 0, 0]  & [0, 0, 0] & [0, 0, 1] & [0, 0, -1] & [0, 0, 1] & [0, 0, -1] \\
          & $\mathrm{Ag_6(b)}$  & [0, 0, 1] & [0, 0, 1] & [0, 0, -1] & [0, 0, -1] &  [0, 0, 0] &  [0, 0, 0] &  [0, 0, 0] &  [0, 0, 0]\\
\end{tabular}
\end{ruledtabular}
\end{table*}

\section{Crystal structure parameters of  \ce{Sm2MnMnMn_{4-x}Ti_xO_{12}}}

We give the crystal structure parameters of \ce{Sm2MnMnMn_{4-x}Ti_xO_{12}} for $x = 1$, $x = 2$ and $x = 3$ in Tables \ref{TAB::xyz_bvs_x=1}, \ref{TAB::xyz_bvs_x=2} and \ref{TAB::xyz_bvs_x=3} respectively.

\begin{table*}[h]
\caption{\label{TAB::xyz_bvs_x=1}Crystal structure parameters of \ce{Sm_2Mn(Mn_{0.84}Sm_{0.16})Mn3TiO12} ($Z=2$, space group $P4_2/nmc$) refined at 80 K. The lattice parameters were determined to be $a =  7.4068(1) ~ \mathrm{\AA}$ and $c = 7.9275(2) ~ \mathrm{\AA}$. Excellent reliability parameters of $R = 2.72\%$, $wR = 2.90\%$, $R_\mathrm{Bragg} = 4.06\%$ were achieved in the refinement. Bond valence sums (BVS) were calculated using the bond valence parameters,  $R_0$(Sm$^{3+}$) = 2.01(1), $R_0$(Mn$^{3+}$) = 1.76(1), $R_0$(Mn$^{2+}$) = 1.79(1), $R_0$(Ti$^{4+}$) = 1.82(1), B = 0.37 \cite{1991Brese}. N.B the Wyckoff position of the Mn1 ions is listed for the split site with an occupation of 0.5, and the high symmetry position is given in square brackets with an occupation of 1. }
\begin{ruledtabular}
\begin{tabular}{c | c c c c c c c}
Ion & Site & Sym. & $x$ & $y$ & $z$ & $U_\mathrm{iso}$ $(\mathrm{\AA}^2)$ & B.v.s. ($|e|$)\\
\hline
Sm1     & $4d$ & $2mm.$      & 0.25      & 0.25       & 0.211(3)   & 0.010(9)  & 3.2(2) \\
Mn1     & $4c [2a]$ & $2mm.$      & 0.75      & 0.25       & 0.720(1)   & 0.009(4)  & 2.34(4) \\
Mn2     & $2b$ & $\bar{4}m2$ & 0.75      & 0.25       & 0.250      & 0.008(5)  & 1.74(4) \\
Mn3/Ti1 & $8e$ & $\bar{1}$   & 0         & 0          & 0          & 0.023(2)  & 3.7(1)/3.23(9) \\
O1      & $8g$ & $.m.$       & 0.25      & 0.0597(4)  & -0.0370(3) & 0.028(2)  & - \\
O2      & $8g$ & $.m.$       & 0.25      & 0.5356(4)  & 0.5770(4)  & 0.028(2)  & - \\
O3      & $8f$ & $..2$       & 0.4355(2) & -0.4355(2) & 0.25      & 0.029(2)  & - \\
\end{tabular}
\end{ruledtabular}
\end{table*}

\begin{table*}[h]
\caption{\label{TAB::xyz_bvs_x=2}Crystal structure parameters of \ce{Sm_2Mn(Mn_{0.93}Sm_{0.07})Mn2Ti2O12} ($Z=2$, space group $P4_2/nmc$) refined at 90 K. The lattice parameters were determined to be $a = 7.5506(1) ~ \mathrm{\AA}$ and $c = 7.7273(3) ~ \mathrm{\AA}$. Excellent reliability parameters of $R = 2.71\%$, $wR = 2.87\%$, $R_\mathrm{Bragg} = 5.06\%$ were achieved in the refinement. Bond valence sums (BVS) were calculated using the bond valence parameters, $R_0$(Sm$^{3+}$) = 2.01(1), $R_0$(Mn$^{2+}$) = 1.79(1), $R_0$(Mn$^{3+}$) = 1.76(1), $R_0$(Ti$^{4+}$) = 1.82(1), B = 0.37 \cite{1991Brese}. N.B the Wyckoff position of the Mn1 ions is listed for the split site with an occupation of 0.5, and the high symmetry position is given in square brackets with an occupation of 1. }
\begin{ruledtabular}
\begin{tabular}{c | c c c c c c c}
Ion & Site & Sym. & $x$ & $y$ & $z$ & $U_\mathrm{iso}$ $(\mathrm{\AA}^2)$ & B.v.s. ($|e|$)\\
\hline
Sm1     & $4d$ & $2mm.$      & 0.25      & 0.25        & 0.209(3)   & 0.022(4)  & 3.2(2) \\
Mn1     & $4c [2a]$ & $2mm.$      & 0.75      & 0.25        & 0.789(1)   & 0.022(4)  & 1.78(5) \\
Mn2     & $2b$ & $\bar{4}m2$ & 0.75      & 0.25        & 0.250      & 0.022(4)  & 1.71(4) \\
Mn3/Ti1 & $8e$ & $\bar{1}$   & 0         & 0           & 0          & 0.014(2)  & 3.7(1)/ 3.22(9) \\
O1      & $8g$ & $.m.$       & 0.25      & 0.0611(4)    & -0.0343(3) & 0.019(2)  & - \\
O2      & $8g$ & $.m.$       & 0.25      & 0.5406(4)   & 0.5705(4)  & 0.022(2)  & - \\
O3      & $8f$ & $..2$       & 0.4436(2) & -0.4436(2) & 0.25       & 0.023(2)  & - \\
\end{tabular}
\end{ruledtabular}
\end{table*}

\begin{table*}[h]
\caption{\label{TAB::xyz_bvs_x=3}Crystal structure parameters of \ce{Sm2MnMnMnTi3O12} ($Z=2$, space group $P4_2/nmc$) refined at 80 K. The lattice parameters were determined to be $a =  7.6290(1) ~ \mathrm{\AA}$ and $c = 7.6932(3) ~ \mathrm{\AA}$. Excellent reliability parameters of $R = 2.03\%$, $wR = 2.20\%$, $R_\mathrm{Bragg} = 4.21\%$ were achieved in the refinement. Bond valence sums (BVS) were calculated using the bond valence parameters,$R_0$(Sm$^{3+}$) = 2.01(1), $R_0$(Mn$^{2+}$) = 1.79(1), $R_0$(Mn$^{3+}$) = 1.76(1), $R_0$(Ti$^{4+}$) = 1.82(1), B = 0.37 \cite{1991Brese}. N.B the Wyckoff position of the Mn1 ions is listed for the split site with an occupation of 0.5, and the high symmetry position is given in brackets with an occupation of 1. }
\begin{ruledtabular}
\begin{tabular}{c | c c c c c c c}
Ion & Site & Sym. & $x$ & $y$ & $z$ & $U_\mathrm{iso}$ $(\mathrm{\AA}^2)$ & B.v.s. ($|e|$)\\
\hline
Sm1     & $4d$ & $2mm.$      & 0.25      & 0.25        & 0.214(3)   & 0.023(3)  & 3.0(2) \\
Mn1     & $4c [2a]$ & $2mm.$ & 0.75      & 0.25        & 0.797(1)   & 0.023(3)  & 1.58(2) \\
Mn2     & $2b$ & $\bar{4}m2$ & 0.75      & 0.25        & 0.250      & 0.023(3)  & 1.81(2) \\
Mn3/Ti1 & $8e$ & $\bar{1}$   & 0         & 0           & 0          & 0.022(2)  & 3.6(1)/3.36(9) \\
O1      & $8g$ & $.m.$       & 0.25      & 0.0641(4)   & -0.0339(4) & 0.027(2)  & - \\
O2      & $8g$ & $.m.$       & 0.25      & 0.5441(4)   & 0.5719(3)  & 0.024(2)  & - \\
O3      & $8f$ & $..2$       & 0.4448(3) & -0.4448(3)  & 0.25       & 0.028(2)  & - \\
\end{tabular}
\end{ruledtabular}
\end{table*}
\end{document}